\documentclass[floatfix,aps,pra,twocolumn,showpacs,superscriptaddress,10pt]{revtex4-2}

\usepackage[english]{babel}
\usepackage{graphicx}
\usepackage{dcolumn}
\usepackage{bm}
\usepackage{subcaption}
\usepackage{float}
\usepackage{hyperref}
\usepackage{ragged2e}
\usepackage{amsmath}
\usepackage{xcolor}
\usepackage{mathtools}
\usepackage{amsfonts}
\usepackage{booktabs}
\usepackage{braket}

\usepackage[most]{tcolorbox}
\usepackage{varwidth}
\usepackage{ragged2e}
\usepackage{xcolor}

\tcbset{
  mechanismpanel/.style={
    enhanced,
    breakable,
    colback=white,
    colframe=teal!60!black,
    colbacktitle=teal!70!black,
    coltitle=white,
    fonttitle=\bfseries,
    boxrule=0.8pt,
    arc=2mm,
    left=2mm,
    right=2mm,
    top=1mm,
    bottom=1mm,
    title filled,
  },
  jsonpanel/.style={
    enhanced,
    breakable,
    colback=black!2,
    colframe=black!20,
    boxrule=0.4pt,
    arc=1mm,
    left=2mm,
    right=2mm,
    top=1mm,
    bottom=1mm,
  },
  verifypanel/.style={
    enhanced,
    breakable,
    colback=blue!2,
    colframe=blue!35!black,
    boxrule=0.5pt,
    arc=1mm,
    left=2mm,
    right=2mm,
    top=1mm,
    bottom=1mm,
  }
}

\newtcblisting[auto counter]{interpretjson}[2][]{
  enhanced,
  breakable,
  listing only,
  listing engine=listings,
  colback=black!2,
  colframe=teal!65!black,
  colbacktitle=teal!72!black,
  coltitle=white,
  fonttitle=\bfseries\small,
  title={#2},
  boxrule=0.7pt,
  arc=1.4mm,
  left=2mm, right=2mm, top=1.2mm, bottom=1.2mm,
  before skip=8pt, after skip=3pt,
  listing options={
    basicstyle=\ttfamily\scriptsize,
    breaklines=true,
    breakatwhitespace=false,
    columns=fullflexible,
    keepspaces=true,
    showstringspaces=false,
    upquote=false,
    tabsize=2,
    xleftmargin=0pt,
    frame=none
  },
  #1
}

\newtcblisting{promptbox}[1]{
    enhanced,
    breakable,
    listing only,
    listing engine=listings,
    colback=black!2,
    colframe=teal!60!black,
    colbacktitle=teal!70!black,
    coltitle=white,
    fonttitle=\bfseries\small,
    title={#1},
    boxrule=0.8pt,
    arc=2mm,
    left=2mm,
    right=2mm,
    top=2mm,
    bottom=2mm,
    before skip=8pt,
    after skip=8pt,
    listing options={
        basicstyle=\ttfamily\footnotesize,
        breaklines=true,
        breakatwhitespace=false,
        columns=fullflexible,
        keepspaces=true,
        showstringspaces=false,
        tabsize=2,
        frame=none
    }
}

\begin{document}

\title{Programming Quantum Circuit Parameters with Large Language Models}

\author{Farias, T.S.}
\email{tiago.farias@ufscar.br}
\affiliation{University of Innsbruck, Department of Physics, Technikerstrasse 25, 6020 Innsbruck, Austria}
\affiliation{Federal University of São Carlos, Department of Computer Science, Rod. Washington Luís, km 235, Jardim Guanabara, São Carlos - SP, 13565-905, Brazil}

\author{Porto, C.M.}
\affiliation{Federal University of São Carlos, Department of Computer Science, Rod. Washington Luís, km 235, Jardim Guanabara, São Carlos - SP, 13565-905, Brazil}

\author{Fonseca, M. S.}
\affiliation{Federal University of São Carlos, Department of Computer Science, Rod. Washington Luís, km 235, Jardim Guanabara, São Carlos - SP, 13565-905, Brazil}

\author{Carpio, N. A. C.}
\affiliation{Federal University of São Carlos, Department of Computer Science, Rod. Washington Luís, km 235, Jardim Guanabara, São Carlos - SP, 13565-905, Brazil}

\author{De Oliveira, A. A.}
\affiliation{Federal University of São Carlos, Department of Computer Science, Rod. Washington Luís, km 235, Jardim Guanabara, São Carlos - SP, 13565-905, Brazil}

\author{Villas-Boas, C. J}
\affiliation{Federal University of São Carlos, Department of Computer Science, Rod. Washington Luís, km 235, Jardim Guanabara, São Carlos - SP, 13565-905, Brazil}

\begin{abstract}
Large language models (LLMs) have recently demonstrated
the ability to generate parameters for variational quantum
circuits without conventional optimization. However,
the conditions under which this approach is effective
remain unclear. Here, we investigate a frozen,
general-purpose LLM that generates complete circuit
parameter vectors from symbolic circuit and Hamiltonian
descriptions, with optional iterative energy feedback.
We evaluate the approach on general random Hamiltonians,
structured parent Hamiltonians, and molecular systems.
For general Hamiltonians, LLM-generated parameters
exhibit larger energy errors than conventional Adam
optimization, while additional feedback generally provides
limited improvement. In contrast, for structured
Hamiltonians, the LLM can generate promising low-energy
parameter configurations, suggesting that identifiable
physical and mathematical structure facilitates parameter
inference. We also show that direct LLM predictions of
minimum energies can violate the Hamiltonian spectral
lower bound, whereas generated circuit parameters allow
independent evaluation of physically realizable energy
expectation values. Finally, we investigate blind circuit
interpretation and find that the LLM can predict
measurement outcomes even when it fails to identify
the underlying quantum algorithm.
These findings highlight the role of problem structure,
the limitations of iterative LLM optimization, and the
importance of independently verifiable predictions in
quantum parameter programming.
\end{abstract}

\maketitle

\onecolumngrid

\section{Introduction}
\label{sec:introduction}

Variational quantum algorithms (VQAs) combine
parameterized quantum circuits with classical
optimization to address computational problems
in quantum simulation, optimization, and machine
learning \cite{cerezo_variational_2021,
peruzzo_variational_2014}. In this framework,
a quantum processor prepares a state using
a set of trainable circuit parameters, while
a classical optimizer updates these parameters
based on evaluations of an objective function.
The variational quantum eigensolver (VQE),
for example, seeks to minimize the expectation
value of a Hamiltonian to approximate its
ground-state energy. Although this hybrid
approach can accommodate the constraints of
current quantum devices, its practical
performance depends on both the quality of
the parameterized circuit and the efficiency
of the optimization procedure.

A central challenge is the number of quantum
circuit evaluations required during optimization.
When the objective function is estimated on
quantum hardware, each evaluation involves
circuit executions and measurements, potentially
across multiple Pauli terms of the Hamiltonian.
Gradient-based approaches introduce additional
evaluations to estimate derivatives. For example,
the parameter-shift rule commonly requires two
shifted expectation evaluations per independently
parameterized rotation \cite{PhysRevA.99.032331}.
As the number of parameters grows, the measurement
demands can become substantial, while statistical
fluctuations and unfavorable optimization
landscapes may further complicate convergence
\cite{mcclean_barren_2018}. Consequently,
reducing the number of quantum objective
evaluations needed to identify useful circuit
parameters is an important consideration in
variational quantum optimization.

Neural networks have been investigated as a
means of learning optimization strategies rather
than relying exclusively on predefined update
rules. Andrychowicz et al.~\cite{andrychowicz2016}
introduced a learning-to-optimize framework in
which recurrent neural networks learn parameter
updates across optimization tasks. Subsequent
studies examined the scalability, generalization,
and computational trade-offs of learned
optimizers \cite{Wichrowska2017,
pmlr-v199-metz22a,moudgil2025celo}.
Unlike conventional optimizers with fixed
update procedures, these approaches acquire
optimization behavior through training on
previous tasks. Their application to new
problems therefore generally depends on
specialized meta-training.
Recent work has additionally examined reactive optimization
rules that respond to local optimization information
\cite{bai_reactive_2026}. Such methods provide a useful
point of comparison for distinguishing inference-time
parameter generation from learned update mechanisms.

Related approaches have been developed for
parameterized quantum circuits. Verdon et
al.~\cite{verdon2019learning} trained classical
recurrent neural networks to propose useful
parameters for variational quantum algorithms,
including QAOA and VQE. Khairy et
al.~\cite{khairy2020learning} investigated
learning-based strategies for optimizing
variational quantum circuits in combinatorial
problems, while Friedrich and
Maziero~\cite{friedrich_learning_2025}
combined neural optimization with evolutionary
strategies. More recently, Zou et
al.~\cite{zou_generative_2025} introduced
Flow-VQE, which learns a generative distribution
of low-energy circuit parameters through
energy-based preference training. These studies
show that learned parameter generation can
reduce optimization demands, although the
resulting models depend on dedicated training
procedures.
Other relevant directions include meta-learning across
quantum optimization tasks \cite{wilson_optimizing_2021},
learning quantum optimization strategies with quantum
neural networks \cite{11250167}, and parameter transfer
between QAOA problems using graph structure or symmetries
\cite{falla_graph_2024,PhysRevA.111.022418}.
Parameter initialization can itself affect trainability,
as investigated through reduced-domain initialization
\cite{PhysRevApplied.22.054005}.
These studies motivate comparing the Language Model Optimizer (LMO)  not only
against generic numerical optimization but also against
structure-aware initialization and transfer strategies.

Large language models have also emerged
as general-purpose optimization agents.
Yang et al.~\cite{yang_large_2024} introduced
Optimization by Prompting, in which
an LLM proposes candidate solutions using
previously evaluated candidates and their
objective values. Huang et
al.~\cite{huang_exploring_2024} subsequently
investigated the capabilities and limitations
of LLMs for numerical black-box optimization.
These approaches motivate the use of language
models as optimization agents, but objective
feedback remains an important source of
problem-specific information during the
search process. A broader survey places
LLM-based optimization within classical
and evolutionary optimization methods
\cite{HUANG2024101663}, while SemanticOpt
explicitly investigates semantic information
in black-box optimization \cite{meindl_semanticopt:_2026}.
Beyond proposing candidate solutions,
LLMs have also been investigated for direct
parameter generation. Wang et al.~\cite{wang_neurogen:_2025} explored
neural-network parameter generation using
LLMs trained on network checkpoints and
task-specific descriptions.

A related and emerging direction is
Weight-Space Programming, in which
language models directly inspect, generate,
or modify model parameters.
Takehi et al.~\cite{takehi2026weightspace}
presented preliminary evidence that a
general-purpose LLM can interpret and edit
neural-network weights directly, without
conventional gradient-based training or
a separately trained optimizer.
Unlike approaches that learn parameter-update
rules, this perspective involves operating
directly on parameter values, potentially
exploiting structural information about
the model and its intended behavior.
We note that, in parallel with the preparation
of this work, Kundu et al.~\cite{kundu2026}
independently investigated direct parameter
generation for variational quantum circuits
using general-purpose language models.
Their study considered VQE, QAOA, and quantum
classification, emphasizing parameter quality
and initialization for classical optimization.
While both works investigate parameter
generation without task-specific training,
our study additionally examines iterative
energy feedback, performance across structured
and general Hamiltonians, the physical
verifiability of energy predictions, and
blind quantum circuit interpretation.

\begin{figure}[t] 
    \centering
    \begin{subfigure}[c]{0.365\textwidth}
        \centering
        \includegraphics[width=\linewidth,trim=0 0 0 0,clip]
        {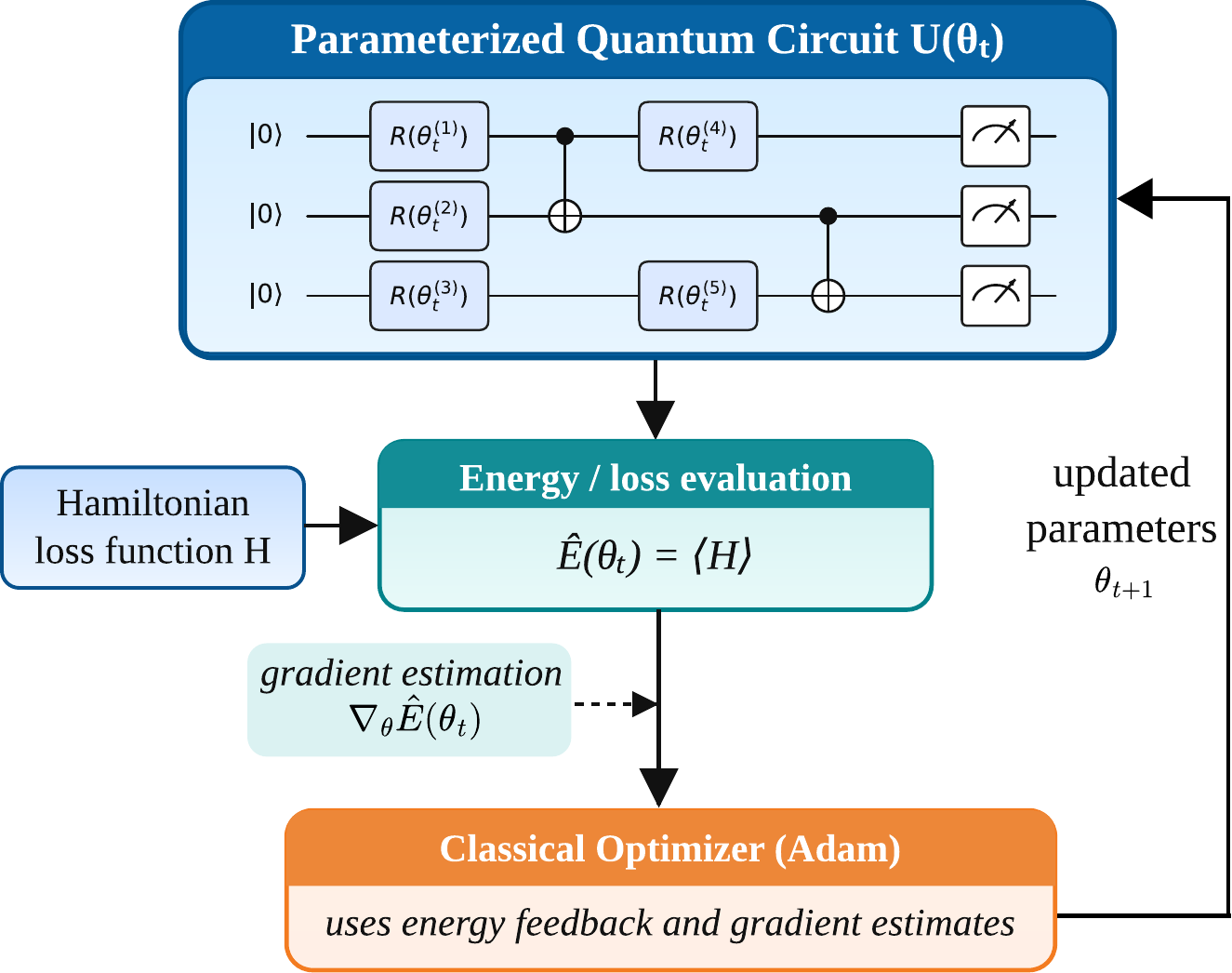}
        \caption{}
    \end{subfigure}
    \hfill
    \begin{subfigure}[c]{0.615\textwidth}
        \centering
        \includegraphics[width=\linewidth,trim=0 0 0 0,clip]
        {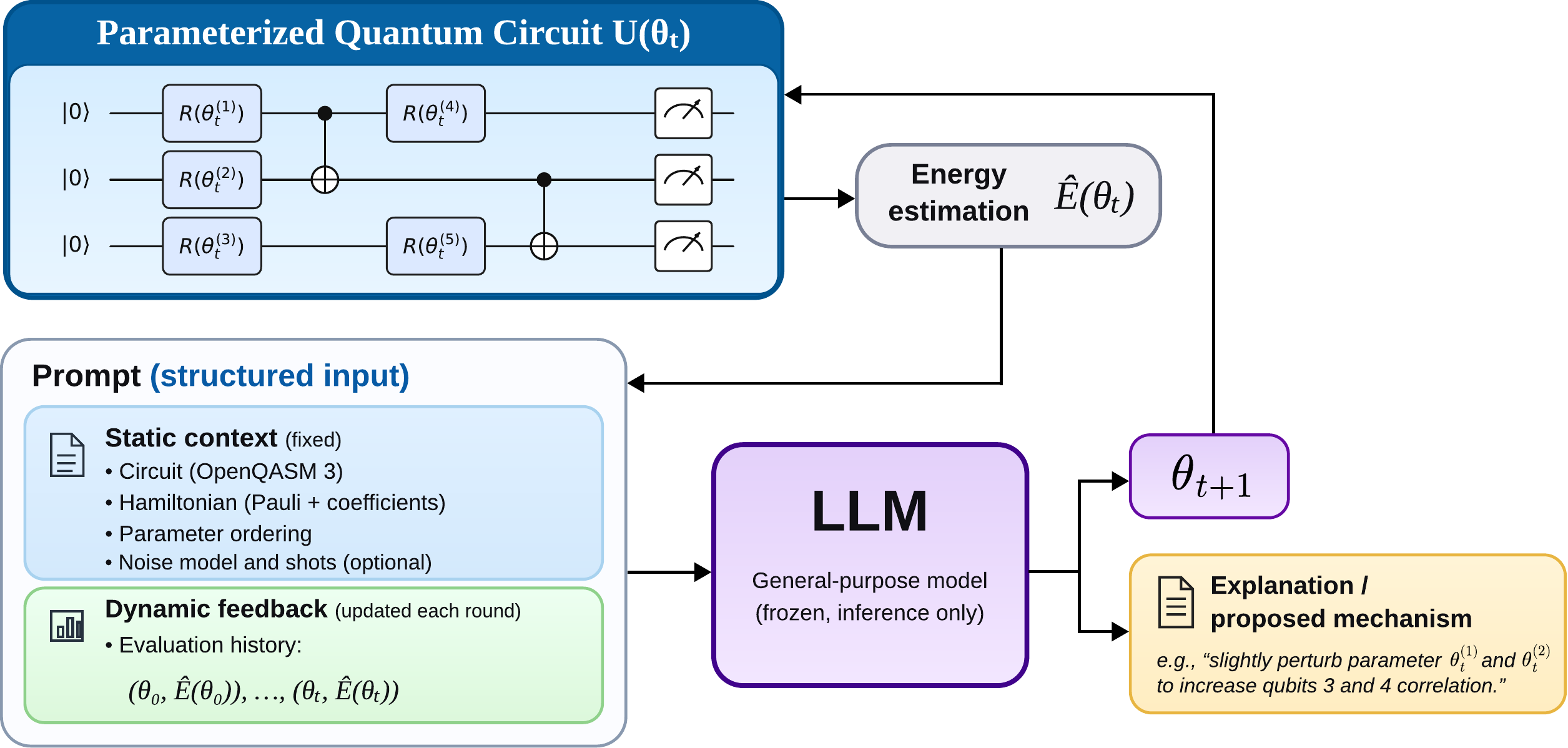}
        \caption{}
    \end{subfigure}

    \caption{\justifying Comparison of conventional
    gradient-based variational optimization and language
    model optimizer.
    (a) A classical optimizer updates circuit parameters
    using quantum objective evaluations and parameter-shift
    gradient estimates.
    (b) A frozen large language model receives a symbolic
    description of the parameterized circuit and Hamiltonian
    and proposes a complete parameter vector without initial
    objective feedback. Subsequent proposals may incorporate
    previously evaluated parameters and their measured
    energies. Both approaches use the same parameterized
    quantum circuit; they differ in how candidate parameters
    are generated.}
    \label{fig:optimizers}
\end{figure}

In quantum computing, language models have
also been investigated for circuit generation
and parameter initialization. Jern et
al.~\cite{Jern_2025} introduced Agent-Q,
which uses fine-tuned LLMs to generate
parameterized quantum circuits. Zhuang and
Guan~\cite{zhuang-guan-2026-large}
proposed AdaInit, which uses LLM-generated
initial parameters and gradient feedback
to improve the trainability of quantum
neural networks. Macarone-Palmieri and
Lo Franco~\cite{macaronepalmieri2026}
investigated iterative circuit synthesis
using simulation feedback. These studies demonstrate the potential of LLMs for quantum circuit generation, parameter initialization, and iterative optimization. The present work focuses specifically on generating complete parameter vectors for a fixed circuit directly from its symbolic representation and target Hamiltonian, without task-specific model training or initial objective-function feedback.
Additional work explores LLM-assisted QAOA circuit
generation \cite{tyagin_qaoa-gpt:_2025}, autonomous
variational circuit design \cite{sakka_llm_2026}, and
LLM-generated circuits and initialization designed to
address trainability \cite{nguyen_bridg-q:_2026}.
These circuit-design tasks differ from the present setting,
where the circuit architecture remains fixed and the LMO
returns its numerical parameters.

Recent evidence that general-purpose LLMs can
directly generate parameters for variational
quantum circuits motivates further investigation
of the conditions under which this capability
is effective. In particular, it remains important
to understand how parameter generation depends
on the structure of the circuit and Hamiltonian,
and whether limited objective-function feedback
can improve subsequent proposals. A quantum
circuit specifies the order and action of its
gates, while the Hamiltonian defines the energy
to be minimized. Together, these descriptions
provide problem-specific information that may
guide parameter generation independently of
previously measured objective values.

In this work, we investigate a
Language Model Optimizer (LMO) that
generates numerical parameters for fixed
quantum circuits, as illustrated in
Fig.~\ref{fig:optimizers}. The approach
uses a frozen, general-purpose LLM that
receives the parameterized circuit in
OpenQASM 3 \cite{cross2022openqasm3} and the Hamiltonian as a
Pauli decomposition. Without task-specific
training or previous objective evaluations
for the given instance, the model proposes
a complete parameter vector. The resulting
circuit is then evaluated, and subsequent
proposals may incorporate previously tested
parameters and their measured energies.
This separates parameter generation from
quantum evaluation and permits symbolic
inference to precede measurement-based
optimization.

This distinction is particularly relevant
when quantum objective evaluations are
expensive. The LMO generates its initial
parameter vector without receiving any
prior energy evaluations. The proposed
parameters can then be evaluated through
a single Hamiltonian expectation-value
estimation to assess their quality.
Depending on the experimental protocol,
the resulting energy may be returned to
the model as feedback for subsequent
proposals or used solely for independent
evaluation. The use of iterative feedback
is specified separately for each experiment.
However, generating parameters without
initial objective feedback does not remove
the need for experimental validation,
nor does it imply that the proposed
parameters are optimal for arbitrary
Hamiltonians. Moreover, the cost of
LLM inference must be distinguished from
the number of quantum objective evaluations.
The present approach therefore investigates
a different information-resource trade-off
rather than establishing a general
computational advantage over classical
optimization.

We evaluate the LMO across general random
Hamiltonians, structured parent Hamiltonians
with known ground-state energies, and molecular
systems. These benchmarks allow us to investigate
the quality of initial parameter proposals,
the influence of problem structure, the effect
of iterative energy feedback, and the number
of objective evaluations required relative
to conventional optimization. We also examine
direct LLM predictions of minimum energies,
distinguishing numerical estimates from
physically realizable and independently
verifiable parameter configurations.
Beyond optimization, we investigate the ability
of the same general-purpose model to interpret
quantum circuits from their symbolic
representations, distinguishing algorithm
identification from functional output prediction.
Together, these experiments characterize
the capabilities and limitations of
inference-time quantum parameter programming
across different problem settings.

\section{Language Model-Based Quantum Circuit Analysis}

We investigate two complementary capabilities of a
frozen, general-purpose large language model:
circuit optimization and circuit interpretability.
The first examines whether the model can generate
numerical parameters that minimize the energy
expectation value of a fixed quantum circuit.
The second evaluates its ability to infer the
computational behavior and possible purpose of
quantum circuits directly from their symbolic
representations, without executing them.

\subsection{Circuit optimization}

We consider the optimization of a fixed parameterized quantum circuit $U(\boldsymbol{\theta})$ acting on an $n$-qubit initial state $\ket{0}^{\otimes n}$, where $\boldsymbol{\theta}\in\mathbb{R}^{p}$ denotes the vector of $p$ adjustable parameters. The prepared state and its energy expectation value are defined as
\begin{equation}
    \ket{\psi(\boldsymbol{\theta})}
    =
    U(\boldsymbol{\theta})\ket{0}^{\otimes n},
    \qquad
    E(\boldsymbol{\theta})
    =
    \bra{\psi(\boldsymbol{\theta})}
    H
    \ket{\psi(\boldsymbol{\theta})}.
\end{equation}

For a Hamiltonian expressed as a linear combination of Pauli
operators,
\begin{equation}
    H=\sum_{j=1}^{M}c_jP_j,
    \qquad
    E(\boldsymbol{\theta})
    =
    \sum_{j=1}^{M}
    c_j\langle P_j\rangle_{\boldsymbol{\theta}},
\end{equation}
where $c_j\in \mathbb{R}$ and
$P_j\in\{I,X,Y,Z\}^{\otimes n}$. The objective is to find a parameter vector that minimizes $E(\boldsymbol{\theta})$. On quantum hardware, the energy is estimated from finite measurement samples, yielding an estimator $\widehat{E}(\boldsymbol{\theta})$.

Motivated by the weight-space programming
perspective~\cite{takehi2026weightspace},
the LMO treats circuit parameters as
numerical quantities that can be directly
inferred and modified by a language model,
rather than exclusively updated through
conventional optimization rules. We employ
a frozen, general-purpose LLM to propose
complete parameter vectors from a symbolic
description of the parameterized circuit
in OpenQASM 3 and the Hamiltonian specified
by its Pauli operators and coefficients. The same system instructions are used across all benchmark instances. They specify the energy-minimization objective, the Pauli-label ordering convention, and the required output format. The LMO is not provided with the hidden generating parameters, known ground-state energy, target state, classical optimization trajectories, or numerical statevectors. No task-specific training or parameter updates are performed on the language model. Let $\mathcal{C}$ denote the parameterized OpenQASM 3 description, $\mathcal{H}=\{(c_j,P_j)\}_{j=1}^{M}$ the Hamiltonian representation, and $\mathcal{N}$ any optional disclosed information. Before receiving objective-value feedback, the first proposal is
generated as
\begin{equation}
    \boldsymbol{\theta}_1
    =
    \mathcal{R}_{\mathrm{LMO}}
    \left(
        \mathcal{C},
        \mathcal{H},
        \mathcal{N}
    \right),
    \label{eq:lmo_initial}
\end{equation}
where $\mathcal{R}_{\mathrm{LMO}}$ represents the language-model inference procedure. The proposed vector is evaluated only after generation, so the initial proposal does not depend on quantum objective evaluations performed for that instance. After the initial evaluation, the LMO
may refine its proposals using the
history of previously evaluated parameter
vectors and their corresponding energy
estimates. This iterative use of
objective-value feedback is related
to LLM-based optimization approaches
such as Optimization by Prompting
\cite{yang_large_2024}.
The feedback history is defined as
\begin{equation}
    \mathcal{F}_t
    =
    \left\{
        \left(
            \boldsymbol{\theta}_i,
            \widehat{E}_i
        \right)
    \right\}_{i=1}^{t}.
\end{equation}

Each inference produces one complete numerical parameter vector, rather than an individual coordinate update. The model also returns a textual explanation of its proposed modification, but this explanation is not used to evaluate or accept the candidate. The optimization feedback consists of the numerical energy estimate obtained from the corresponding circuit evaluation. To assess the use of quantum resources, we distinguish Hamiltonian expectation-value evaluations. Each evaluated LMO proposal requires one complete Hamiltonian expectation-value
evaluation to obtain objective feedback,
irrespective of the number of circuit
parameters. For a Hamiltonian containing multiple Pauli terms, a complete evaluation may require
measurements in several bases and repeated circuit executions. In the simulations presented in this work, we count complete energy evaluations without assigning a physical measurement-shot cost.

\subsection{Circuit interpretability}
\label{sec:circuit_interpretability}

Beyond generating circuit parameters, we investigate
whether a general-purpose large language model can
infer the computational behavior of quantum circuits
from their symbolic representations without executing
them. Related studies have investigated the interpretation
of computational models from their internal
representations. Kabra et
al.~\cite{kabra_weight_2026} developed Weight
Oracles, using specialized language models to
infer properties of neural networks directly
from their weights. In quantum computing,
Chen et al.~\cite{chen_symbolic_2026}
investigated symbolic interpretation of Grover
circuits using a domain-adapted language model.
More recently, Mondal~\cite{mondal2026quantumsemeval}
introduced QuantumSemEval, a benchmark evaluating
whether LLMs can recognize semantic equivalence
between structurally different quantum programs.
Related work includes GroverGPT, a language model
focused on quantum search \cite{wang_grovergpt:_2025},
and benchmarks of quantum-computing reasoning
\cite{afane_quantum-audit:_2026} and quantum-code
generation evaluation \cite{seefried_benchmarking_2026}.
Broader perspectives on artificial intelligence for quantum
computing are also available \cite{alexeev_artificial_2025}.
Together, these works motivate evaluating functional
predictions separately from algorithm-label recognition.

Here, we investigate whether a frozen,
general-purpose LLM can infer the computational
behavior of individual quantum circuits under
semantics-preserving changes in representation.
Unlike pairwise equivalence checking, our
protocol evaluates each representation
independently and distinguishes algorithm
identification from functional output prediction.

\begin{figure}[H]
    \centering
    \includegraphics[width=0.8\textwidth]
    {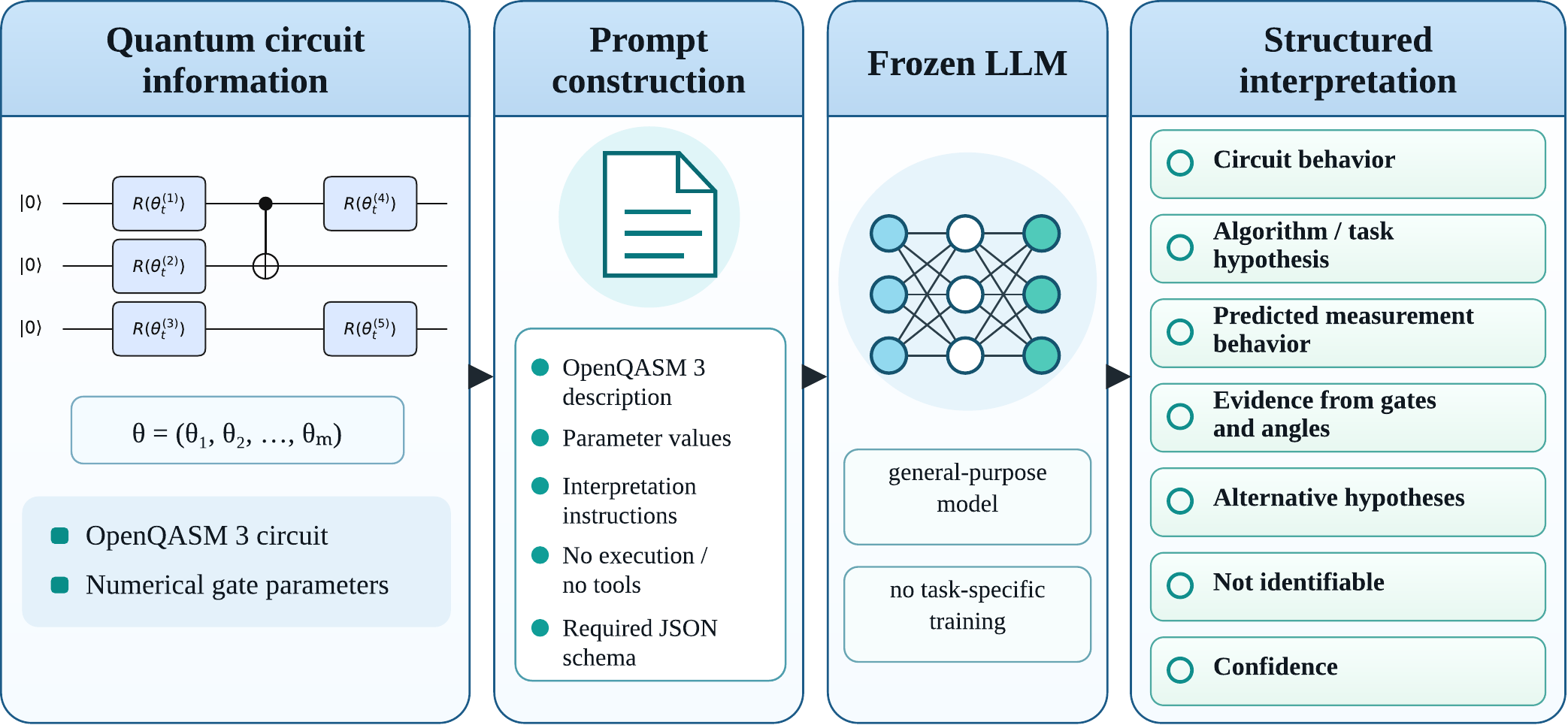}
    \caption{\justifying
        Blind interpretation of quantum circuits using a
        frozen large language model. A circuit and its
        numerical parameters are serialized into OpenQASM 3
        and provided without algorithm-specific metadata.
        The model generates a structured interpretation
        describing the circuit behavior, possible computational
        purpose, expected measurement outcomes, supporting
        gate-level evidence, alternative hypotheses, and
        uncertainty. Its predictions are subsequently
        compared with independently established circuit
        properties.
    }
    \label{fig:circuit_interpretability}
\end{figure}

We formulate this task as blind circuit interpretation.
Given a circuit represented in OpenQASM 3, the frozen LLM
is asked to describe its mathematical behavior, identify
its possible computational purpose, and predict observable
outcomes when these can be determined, as illustrated in
Fig.~\ref{fig:circuit_interpretability}. The input includes
the gate operations, numerical parameters, qubit interactions,
and measurement instructions, but excludes the algorithm
name, expected solution, and benchmark metadata. The model
receives no statevectors, simulated measurements, or
experimental feedback.

We consider three quantum algorithms with distinct
computational behaviors: Grover's search algorithm,
the quantum Fourier transform (QFT), and the
Bernstein-Vazirani algorithm. For each algorithm,
we construct five instances with different problem
parameters or input states. Each instance is expressed
in three representations: logical, using standard
quantum gates; native, obtained by decomposition
into the restricted gate basis
$\{R_Z,R,R_{XX}\}$; and obfuscated, obtained
through randomized local Clifford-frame transformations
followed by native-gate decomposition. The transformations
preserve the implemented computation, which is verified
numerically before inference.

The three representations allow us to investigate whether
the model's interpretation depends on familiar gate
sequences. Logical circuits retain recognizable algorithmic
structures, whereas native decomposition and randomized
frame transformations modify their gate-level descriptions.
Algorithm-specific identifiers and comments are removed
from all representations. The resulting circuit descriptions
are evaluated independently, without providing the model
with their equivalence relationships.

For every circuit, the LLM receives the same instructions
and returns a structured response containing its inferred
circuit behavior, algorithm hypothesis, predicted measurement
behavior, supporting evidence, alternative hypotheses,
and uncertainty. When a specific output bitstring can be
inferred, the model may report that prediction explicitly,
otherwise, it may abstain. No external tools, simulations,
or circuit executions are available to the model during
inference. The ground-truth circuit properties are determined
independently and are not included in the model input.
The different representations preserve the underlying
computation but may differ in circuit depth and description
length. Consequently, this protocol evaluates robustness
under representation changes rather than isolating the
effect of obfuscation alone. Correct interpretation of
transformed circuits does not, by itself, rule out the
use of patterns learned during language-model training.
Further implementation details and evaluation criteria
are provided in Appendix~\ref{app:circuit_interpretability}.

\section{Results}
\label{sec:results}

This section presents the experimental evaluation of the LMO
across parameter optimization, iterative feedback, and circuit
interpretation. For parameter optimization, we consider two
families of problems: general Hamiltonians and structured
Hamiltonians, including parent and molecular Hamiltonians. The results support four main observations:
\begin{itemize}
    \item \textbf{General Hamiltonians.}
    For general Hamiltonians, the LLM struggles to identify
    low-energy parameter configurations and does not
    consistently outperform conventional optimization
    with Adam.

    \item \textbf{Structured Hamiltonians.}
    For Hamiltonians with identifiable physical or mathematical
    structure, such as parent and molecular Hamiltonians,
    the LLM can generate promising parameter configurations,
    suggesting that problem-specific structure may facilitate
    parameter inference. In some molecular experiments,
    successive LMO proposals yield further reductions in energy,
    although the contribution of energy feedback cannot be
    isolated from the effect of generating additional proposals.
    
    \item \textbf{Iterative feedback.}
    Successive LMO proposals generally provide limited
    improvement over the initial parameter configuration,
    although substantial energy reductions are observed
    in some molecular experiments. Since additional
    proposals are generated alongside the incorporation
    of previous energy evaluations, the specific
    contribution of feedback remains undetermined.

    \item \textbf{Circuit interpretability.}
    The LLM can identify certain quantum algorithms from
    their circuit representations, including some obfuscated
    circuits. Even when algorithm identification fails,
    the model can sometimes correctly predict the most
    likely measurement bitstring.
\end{itemize}

All LLM-based experiments reported in this work use
GPT-6 Astra. 
For experiments involving the Adam optimizer \cite{kingma2015adam},
we performed a preliminary learning-rate search
over $\{10^{0}, 10^{-1}, 10^{-2}\}$.
The learning rate yielding the lowest final energy
under a fixed optimization budget was selected
for the subsequent experiments.

\subsection{Circuit optimization}

We evaluate the LMO on three classes of optimization
problems: molecular Hamiltonians, general random
Hamiltonians, and structured parent Hamiltonians.
These problems differ in their physical interpretation
and mathematical structure, allowing us to investigate
how these properties influence the quality of
LLM-generated circuit parameters. For each problem,
we examine the initial parameter proposals, the
effect of iterative energy feedback, and their
performance relative to conventional optimization
with Adam.

\subsubsection{$H_4$ Molecule}
\label{sec:h4_molecule}

\begin{figure}[h]
    \centering
    \includegraphics[width=0.8\textwidth]
    {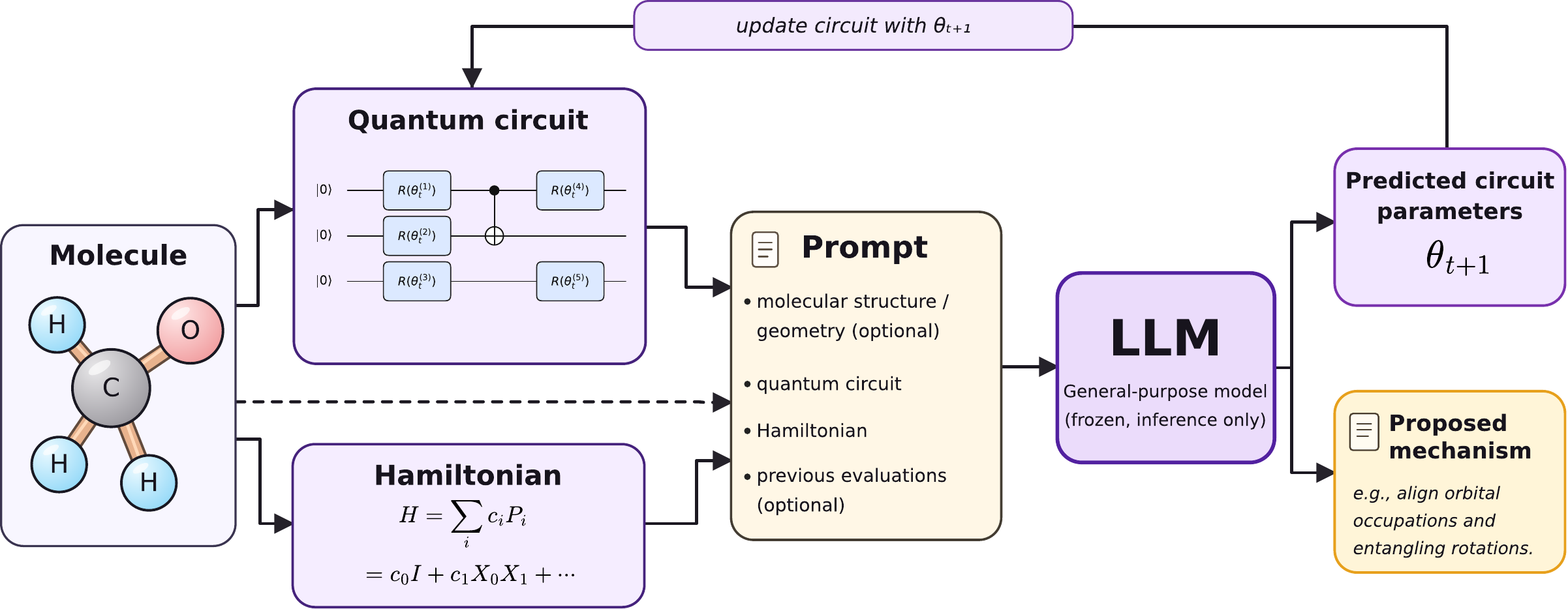}
    \caption{\justifying
    Workflow of language-model-based parameter optimization
    for molecular Hamiltonians. A molecular system is
    represented by a qubit Hamiltonian and a fixed
    parameterized quantum circuit. Their symbolic
    descriptions are incorporated into a prompt, optionally
    supplemented with molecular information and previous
    energy evaluations. A frozen, general-purpose LLM
    generates a complete parameter vector and a textual
    explanation of its proposed mechanism. The predicted
    parameters are assigned to the circuit, which can
    subsequently be evaluated to provide energy feedback
    for further proposals. Solid arrows indicate the
    main information and parameter flow, while the dashed
    arrow represents optional molecular information
    supplied directly to the prompt.
    }
    \label{fig:molecule_workflow}
\end{figure}

We investigate the LMO on the molecular
ground-state energy problem of $H_4$, considering
a fixed planar molecular geometry described using the
STO-3G basis. The system contains four electrons
distributed over four spatial orbitals, without
additional active-space truncation, generating a total of 8 spin-orbital, the same as the number of qubits.
As illustrated in Fig.~\ref{fig:molecule_workflow},
the molecular system is represented by a qubit
Hamiltonian which generates a fixed 
UCCSD\cite{Romero_2019} parameterized quantum
circuit. These representations are provided to
the LMO, which proposes complete parameter
vectors together with a textual explanation
of the proposed mechanism. The generated
parameters are used to instantiate the circuit,
whose energy expectation value is evaluated
and may be incorporated into subsequent proposals.
Details of Hamiltonian construction and the circuit
ansatz are provided in Appendix~\ref{app:molecular_hamiltonians}. The molecular coordinates and additional $H_4$ results are given in Appendix~\ref{app:h4_molecule}.

We consider two prompting strategies. In the
blind configuration, the LMO receives the
parameterized circuit description and the
Hamiltonian as a Pauli decomposition, without
additional chemical information. In the
physics-aware configuration, the same inputs
are supplemented with molecular information,
including the geometry, basis set, electronic
configuration, active space, and reference state,
as indicated by the dashed arrow in
Fig.~\ref{fig:molecule_workflow}.
The standard optimization instructions and
parameter-output format are preserved, with
the physics-aware instructions additionally
allowing the use of chemical context.
Both configurations use the same frozen
language model and circuit architecture.
The corresponding prompts are documented
in Appendix~\ref{app:system_prompts}.

To examine the variability of the generated
parameters, each prompting configuration is
evaluated in three independent runs using the same
initial prompt and language model. Within each run,
the LMO may incorporate previously evaluated
parameter vectors and their corresponding energies
to generate subsequent proposals. We compare the
resulting optimization trajectories with those of
the gradient-based Adam optimizer. Performance is
measured by the energy error relative to the
active-space full configuration interaction (FCI)
reference, expressed as a function of the cumulative
number of Hamiltonian energy evaluations.

\begin{figure}[h]
    \centering
    \includegraphics[width=1.0\textwidth]{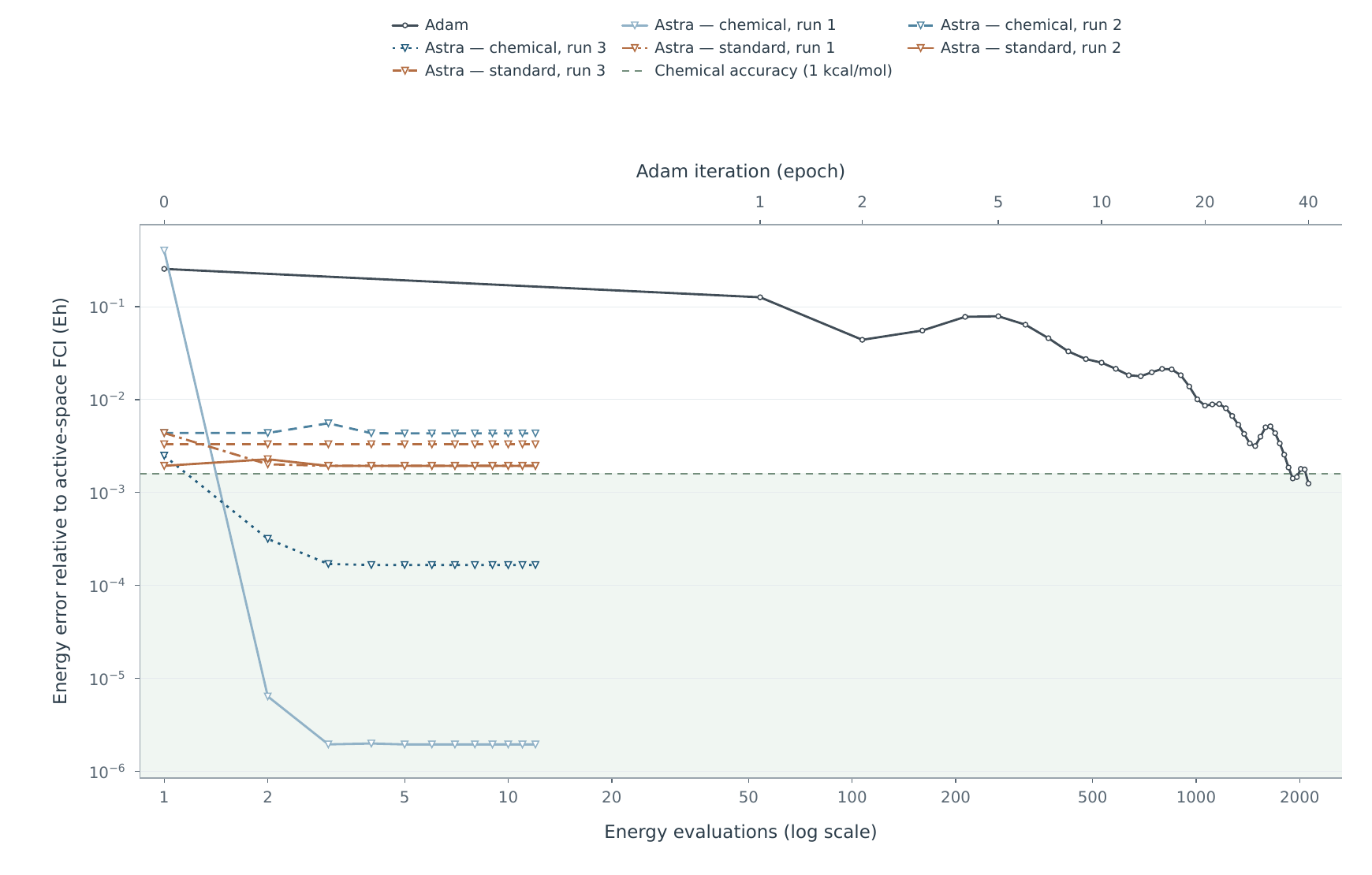}
    \caption{\justifying
    Optimization of the $H_4$ molecular Hamiltonian
    at the standard planar geometry,using Adam and the language model optimizer.
    The LMO is evaluated with a standard blind prompt
    and a physics-aware prompt containing additional
    chemical information, denoted as standard and
    chemical, respectively. Three independent runs
    are shown for each prompting configuration.
    The vertical axis represents the energy error
    relative to the active-space full configuration
    interaction reference, while the horizontal
    axis shows the cumulative number of Hamiltonian
    energy evaluations on a logarithmic scale.
    The dashed horizontal line indicates the
    chemical-accuracy threshold of $1$ kcal/mol.
    The upper horizontal axis indicates the
    corresponding Adam optimization iterations.
    }
    \label{fig:h4_gap}
\end{figure}

Figure~\ref{fig:h4_gap} presents the optimization
trajectories for the standard planar geometry.
The physics-aware prompt produces the lowest
energy errors, with some runs approaching the
reference energy after only a few evaluations.
Several physics-aware trajectories show further
energy reductions across successive proposals,
whereas the blind configurations exhibit
comparatively limited improvement. These
observations suggest that additional molecular
information may facilitate the generation of
effective parameter configurations. However,
the contribution of energy feedback cannot be
isolated from that of generating additional
candidate parameters, and performance varies
across independent runs. Adam follows a more
gradual optimization trajectory, continuing
to reduce the energy error over a substantially
larger evaluation budget. The comparison
concerns Hamiltonian expectation-value
evaluations rather than the total computational
cost of LLM inference.

To probe generalization beyond the standard geometry,
we also evaluate a distorted, nonplanar $H_4$ structure
with modified Hamiltonian coefficients. Its coordinates,
motivation, and optimization trajectories are reported
in Appendix~\ref{app:h4_molecule}.

\subsubsection{Random Hamiltonians}
\label{sec:random_hamiltonians}

We consider a family of randomly generated Hamiltonians
to evaluate the ability of the LMO to identify low-energy
circuit parameters in the absence of an explicitly
structured physical problem. Each Hamiltonian is defined as
\begin{equation}
    H = \sum_{j=1}^{M} c_j P_j,
    \qquad
    P_j \in \{I,X,Y,Z\}^{\otimes n},
\end{equation}
where $M$ is the number of Pauli terms, $n$ is the
number of qubits, and the coefficients are independently
sampled from
\begin{equation}
    c_j \sim \mathcal{N}
    \left(0,\frac{1}{M}\right).
\end{equation}
Each Pauli string contains exactly three non-identity
operators, selected from $\{X,Y,Z\}$ and assigned to
distinct qubit positions. Duplicate strings are excluded,
and each generated Hamiltonian is required to contain
at least one pair of anticommuting Pauli terms.

We investigate systems with $n\in\{4,8,12\}$ qubits,
using a fixed ansatz with $L=n$
entangling layers. Each rotation block applies
$R_X$, $R_Y$, and $R_Z$ gates to every qubit,
followed by a brickwork arrangement of CNOT gates
connecting nearest-neighbor qubits. The entangling
layers alternate between even and odd neighboring
pairs, with an additional rotation block at the end of the circuit. The resulting ansatz contains
$p=3n(L+1)$ trainable parameters. For each system size, we generate five independent
Hamiltonian instances, each containing $M=16$
Pauli terms.

We compare three approaches on the same Hamiltonian
instances and circuit architectures. First, Adam
optimizes randomly initialized circuit parameters
using parameter-shift gradients. Second, the LMO
generates complete parameter vectors from the circuit description and the Hamiltonian
expressed as Pauli operators and coefficients.
Subsequent LMO proposals may incorporate previously
evaluated parameters and their energy expectation
values as feedback. Third, we investigate direct
energy prediction, in which the LLM estimates the
minimum energy attainable by the fixed parameterized
circuit without generating or evaluating a parameter
vector. Unlike Adam and the LMO, this approach does
not prepare a quantum state and therefore does not
produce a variational energy estimate. Consequently,
its predicted energy may fall below the true
Hamiltonian ground-state energy.

All circuit energy expectation values are computed
using noiseless statevector simulation. The exact
Hamiltonian ground-state energy, obtained through
numerical diagonalization, serves as the reference
for calculating the energy gap
$\Delta E = E - E_0$, where $E_0$ is the lowest
eigenvalue of $H$ and $E$ denotes either the
evaluated circuit energy or the directly predicted
minimum energy, depending on the approach.
The reference energy is not disclosed to the LLM.
Unlike the parent-Hamiltonian benchmark, the
random-Hamiltonian construction does not guarantee
that the chosen ansatz can represent the ground
state. Consequently, a nonzero energy gap may
reflect both limitations in circuit expressivity
and incomplete optimization. Since the minimum
energy attainable within the ansatz is not
independently established, these contributions
cannot be distinguished in the present comparison.

\begin{figure}[h]
    \centering
    \begin{subfigure}[c]{0.47\textwidth}
        \centering
        \includegraphics[width=\linewidth]
        {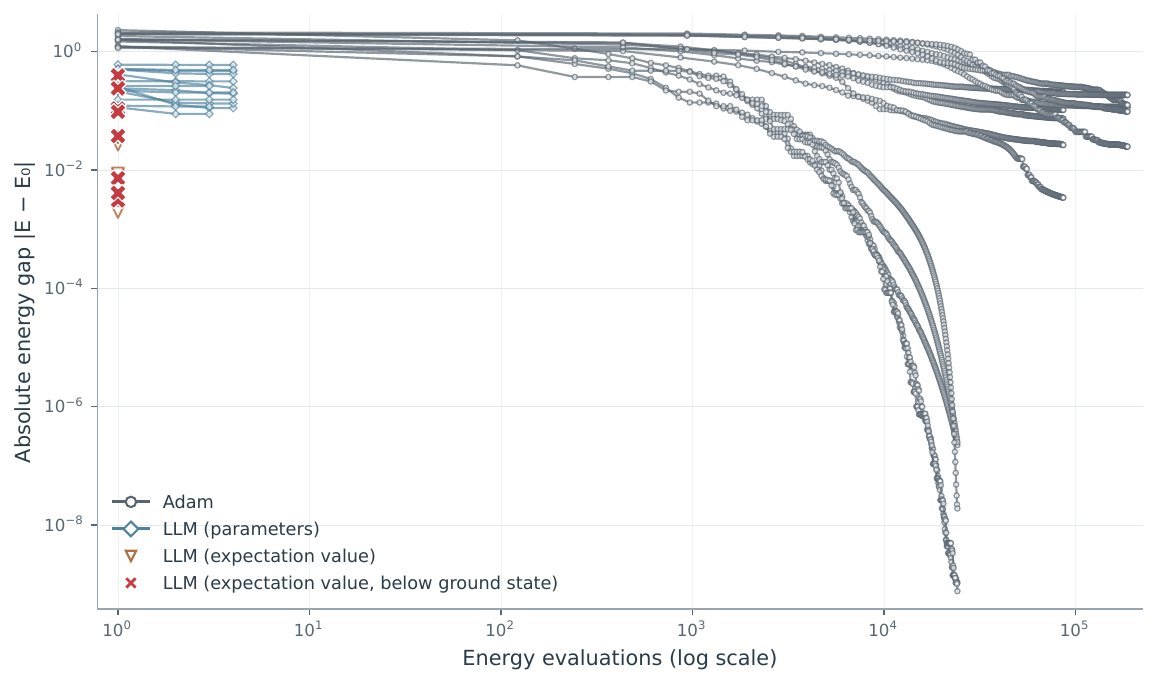}
        \caption{}
    \end{subfigure}
    \hfill
    \begin{subfigure}[c]{0.51\textwidth}
        \centering
        \includegraphics[width=\linewidth]
        {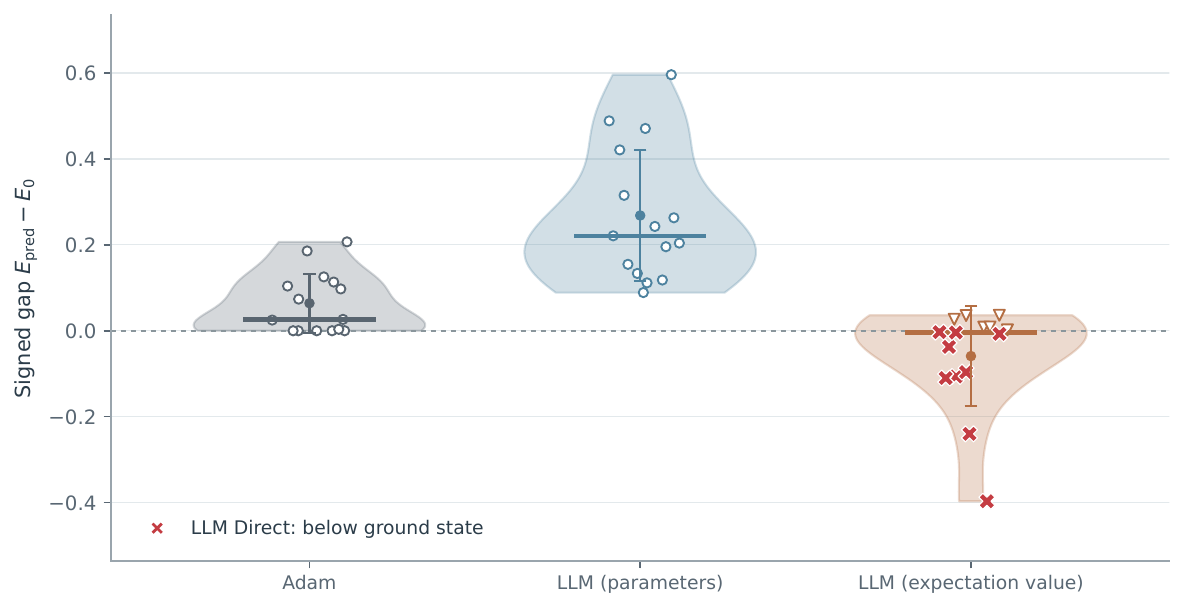}
        \caption{}
    \end{subfigure}
    \caption{\justifying
    Comparison of Adam optimization, LMO parameter generation,
    and direct LLM energy prediction across random Hamiltonian
    instances with different numbers of qubits and circuit layers.
    (a) Best absolute energy gap as a function of the number of
    Hamiltonian expectation-value evaluations. Each trajectory
    corresponds to an individual Hamiltonian instance.
    Direct expectation value LLM predictions are displayed at a nominal position
    of one evaluation for visualization, although they require
    no quantum circuit evaluations.
    (b) Distribution of the signed energy gaps across all
    configurations. Each point represents one Hamiltonian
    instance, and violins summarize the corresponding
    distributions. Red crosses indicate direct LLM predictions
    below the exact ground-state energy, which violate the
    spectral lower bound.}
    \label{fig:random_hamiltonians}
\end{figure}

Figure~\ref{fig:random_hamiltonians}(a) compares the
absolute energy gaps obtained by Adam, the LMO, and
direct LLM energy prediction. Adam progressively reduces
the energy gap through iterative optimization, whereas
the LMO generally produces parameter configurations with
larger gaps and shows limited improvement over subsequent
rounds. This suggests that, for general tested Hamiltonians,
the symbolic circuit and Hamiltonian descriptions alone
provide insufficient information for the LLM to reliably
identify low-energy parameter configurations.

Figure~\ref{fig:random_hamiltonians}(b) presents the
distribution of signed energy gaps across all configurations.
Direct LLM energy prediction frequently produces estimates
below the true ground-state energy, violating the spectral
lower bound. Without prior knowledge of the ground-state
energy or an independently established lower bound, such
violations cannot generally be identified. Moreover, a
directly predicted energy does not provide a quantum state
whose expectation value can be independently evaluated.
In contrast, the LMO generates explicit circuit parameters
that can be used to prepare a quantum state and measure
its energy expectation value, even when the true ground-state
energy is unknown. By the variational principle, this
evaluated energy provides an upper bound on the ground-state
energy, subject to statistical and experimental uncertainties.
Therefore, although LMO-generated parameters do not
necessarily yield lower energy errors, they provide a
physically realizable and independently verifiable energy
estimate, unlike direct numerical predictions.

\subsection{Circuit interpretability}
\label{sec:results_circuit_interpretability}

We evaluate the ability of the frozen LLM to interpret
quantum circuits under three equivalent representations:
logical, native, and obfuscated. The benchmark comprises
Grover's search algorithm \cite{grover1996}, the quantum Fourier transform
(QFT) \cite{shor1994}, and the Bernstein-Vazirani algorithm \cite{bernstein1997}, with five
instances per algorithm and representation, yielding
45 independent interpretations. The model receives only
the anonymized OpenQASM 3 description of each circuit,
without algorithm-specific metadata or access to
circuit execution.

\begin{figure*}[t]
    \centering
    \begin{subfigure}[t]{0.49\textwidth}
        \centering
        \includegraphics[width=\linewidth]
        {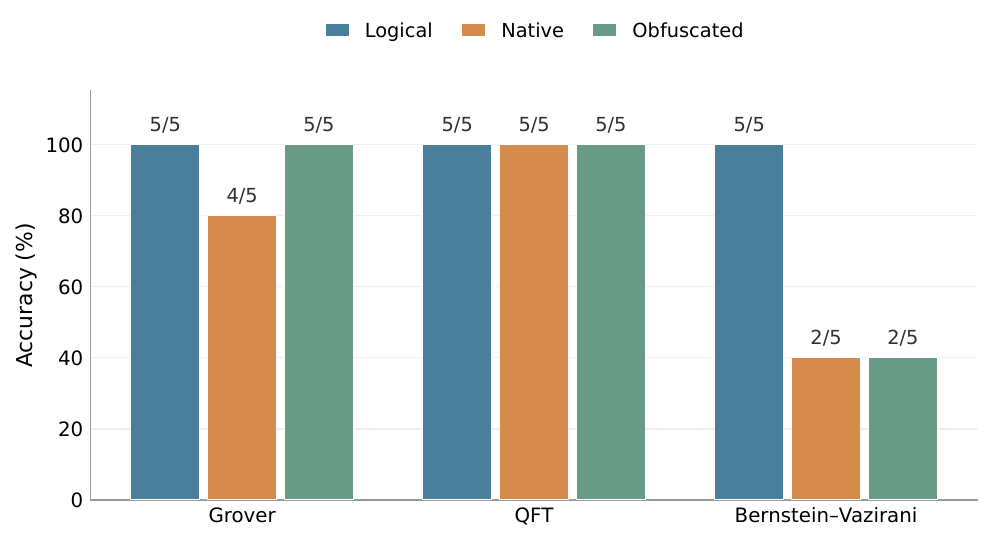}
        \caption{}
        \label{fig:interpretability_identification}
    \end{subfigure}
    \hfill
    \begin{subfigure}[t]{0.49\textwidth}
        \centering
        \includegraphics[width=\linewidth]
        {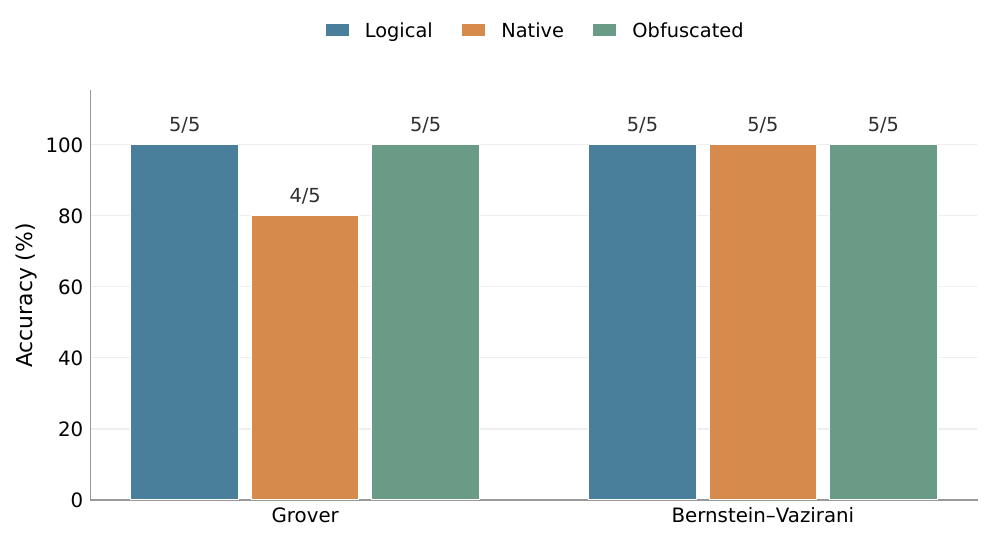}
        \caption{}
        \label{fig:interpretability_output}
    \end{subfigure}
    \caption{\justifying
        Blind quantum circuit interpretation across
        logical, native, and obfuscated representations.
        (a) Accuracy of algorithm identification for
        Grover, QFT, and Bernstein-Vazirani.
        (b) Accuracy of the predicted most-likely
        measurement bitstring for Grover and
        Bernstein--Vazirani.
        Each group contains five instances, and
        labels indicate the number of correct
        predictions. QFT is excluded from (b)
        because a single output bitstring does
        not characterize the Fourier transformation.
    }
    \label{fig:interpretability_results}
\end{figure*}

Figure~\ref{fig:interpretability_results}(a) summarizes
algorithm identification accuracy across the three
representations. The model correctly identifies all
15 algorithms in the logical representation, compared
with 11 of 15 in the native representation and
12 of 15 in the obfuscated representation. Grover's
algorithm is identified in 5/5 logical, 4/5 native,
and 5/5 obfuscated circuits, while QFT is identified
in all 15 cases. The largest reduction occurs for
Bernstein-Vazirani, whose identification accuracy
decreases from 5/5 in the logical representation to
2/5 in both native and obfuscated representations.
These results indicate that algorithm identification
is sensitive to the circuit representation, although
the effect varies across algorithms.

Algorithm identification does not necessarily reflect
whether the model correctly predicts the circuit's
computational output. Figure~\ref{fig:interpretability_results}(b)
reports the accuracy of explicit output bitstring
predictions for Grover and Bernstein-Vazirani.
Across these 30 interpretations, the model predicts
the correct most-likely measurement outcome in
29 cases (96.7\%). Grover reaches 5/5 correct
predictions in the logical representation, 4/5
in the native representation, and 5/5 in the
obfuscated representation. For Bernstein-Vazirani,
the predicted output is correct in all 15 cases,
including those in which the model does not
identify the algorithm by name.

This distinction is particularly evident for
Bernstein-Vazirani. Under native decomposition
and obfuscation, the model correctly identifies
the algorithm in only 2/5 instances per
representation, yet predicts the corresponding
secret bitstring in all five. Thus, a failure
to identify the conventional algorithmic
family does not necessarily imply a failure
to infer the observable computation.
The result also suggests that functional
predictions can remain accurate when familiar
gate patterns are modified by
semantics-preserving transformations. 
For QFT, the model identifies the algorithm
in all 15 cases. However, algorithm recognition
alone does not establish that the model has
correctly characterized the Fourier
transformation. In particular, measurement
probabilities in the computational basis do
not determine the relative phases of the
output state. We therefore distinguish
successful algorithm identification from
the correctness of the model's explanations
of phase relationships and interference,
which are not systematically evaluated here.

The results show that the LLM
can identify familiar quantum algorithms
and frequently predict their computational
outputs under different circuit
representations. The Bernstein-Vazirani
results provide a clear example in which
functional output prediction remains accurate
despite reduced algorithm identification.
Nevertheless, the transformed circuits
also differ in depth and description length,
and the benchmark contains only five
instances per algorithm. Consequently,
these findings demonstrate representation
robustness within the tested circuit
families, but do not establish that the
model's interpretations are independent
of training data patterns or that its
explanations faithfully describe its
internal reasoning process. Additional
implementation details and analysis
are provided in
Appendix~\ref{app:circuit_interpretability}.

\section{Conclusions}
\label{sec:conclusion}

In this work, we investigated the capabilities and
limitations of a frozen, general-purpose large language
model for generating parameters of fixed variational
quantum circuits from symbolic circuit and Hamiltonian
descriptions. The results reveal a clear distinction
between general and structured optimization problems.
For randomly generated Hamiltonians, the LMO struggles
to identify low-energy parameter configurations and
does not consistently outperform conventional Adam
optimization. In contrast, for structured parent and
molecular Hamiltonians, the model can generate
promising parameter configurations without prior
objective-function feedback, suggesting that
identifiable physical and mathematical structure
may facilitate parameter inference.

Iterative feedback generally provides limited
improvement over the initial parameter proposals,
although the physics-aware molecular experiments
demonstrate that additional problem-specific
information and energy feedback can improve
optimization in some cases. We also examined direct
LLM predictions of minimum energies, which can
violate the Hamiltonian spectral lower bound.
Unlike such numerical predictions, LLM-generated
circuit parameters define quantum states whose
energy expectation values can be independently
evaluated, even without knowing the ground-state
energy. This distinction highlights the practical
importance of physically verifiable parameter
proposals.

Beyond parameter optimization, we investigated
blind circuit interpretation under different
symbolic representations. The model successfully
identified several quantum algorithms and
frequently predicted their measurement outcomes,
including cases where algorithm identification
was unsuccessful. These observations suggest that
functional output prediction and algorithm
recognition are distinct capabilities, although
the results do not establish that the model's
explanations faithfully reflect its internal
reasoning process.

Taken together, these findings suggest that
inference-time parameter programming can
complement conventional variational optimization
when the problem admits exploitable structure.
However, the results do not establish a general
computational advantage or demonstrate that the
model performs general quantum-mechanical reasoning.
Its performance may depend on patterns encountered
during pretraining, and the structured Hamiltonians
considered here may admit specialized classical
solution strategies. Furthermore, the cost of
language-model inference must be considered
alongside quantum objective-evaluation requirements.

\section*{Data availability}

The code and data presented in this paper can be found at \url{https://github.com/tiago939/NeuralOptimizer}.

\section*{Acknowledgements}

This work was supported by the São Paulo Research Foundation (FAPESP), under Grants 2025/27502-3, No. 2023/15739-3, No. 2026/22051-6, No. 2025/15490-0, and No. 2022/00209-6, by the National Council for Scientific and Technological Development (CNPq), grants No. 422316/2023-7, No. 140174/2026-5, and No. 302234/2026-8, and by Petrobrás/CENPES, grant No. 2025/00505-2.

\bibliography{references}

\appendix

\section{System Prompts}
\label{app:system_prompts}

This appendix documents the instructions and input formats
used for circuit optimization and blind circuit interpretation.
The system instructions remain fixed within each experimental
protocol, while the user prompts contain instance-specific
information.

\subsection{Blind optimization prompt}
\label{app:blind_prompt}

For circuit optimization, the language model receives a
fixed parameterized circuit expressed in OpenQASM 3 and a
Hamiltonian represented by its Pauli decomposition.
The system instructions require the model to propose a
complete numerical parameter vector that minimizes the
Hamiltonian expectation value, without modifying the
circuit or Hamiltonian. The parameter ordering and Pauli
labeling conventions are explicitly specified.

\begin{promptbox}{Blind optimization - System prompt}
You optimize the expectation value of a supplied Hamiltonian using a fixed parameterized quantum circuit. Choose one complete absolute numerical parameter vector in radians. Do not modify the circuit or Hamiltonian. Infer parameter order from the OpenQASM 3 input declarations in their textual order. Qiskit Pauli labels run from q[n-1] on the left to q[0] on the right. Use only the provided circuit, Hamiltonian, optional noise model, optional measurement specification, and previous angle-energy observations. Return only a JSON object with keys "theta" (numeric array) and "mechanism" (brief, testable reasoning).
\end{promptbox}

The user prompt contains the circuit description and
Hamiltonian coefficients. Depending on the experimental
configuration, additional fields provide information
about the noise model, measurement budget, and previous
parameter evaluations. Its structure is illustrated
below.

\begin{promptbox}{Blind optimization - User prompt structure}
CIRCUIT_OPENQASM_3:
<parameterized OpenQASM 3 circuit>

HAMILTONIAN_PAULI_TERMS:
[
  {
    "pauli": "<Pauli string>",
    "coefficient": <numerical value>
  },
  ...
]

NOISE_MODEL:
<optional JSON noise specification>

MEASUREMENT_SPECIFICATION:
{
  "shots_per_energy_evaluation": <integer>
}

PREVIOUS_EVALUATIONS:
[
  {
    "theta": [<numerical parameters>],
    "energy": <numerical energy>
  },
  ...
]
\end{promptbox}

The first proposal is generated without previous
objective-function evaluations. In subsequent rounds,
the prompt may include the parameter vectors and
corresponding energy estimates obtained from earlier
evaluations. Noise information and measurement
specifications are disclosed only when applicable.
The model returns a JSON object containing the complete
parameter vector and a brief explanation of its proposed
solution. This explanation is recorded but does not
influence the numerical evaluation of the candidate.

\subsection{Physics-aware prompt}
\label{app:physics_aware_prompt}

For molecular problems, the LMO receives the same
circuit and Hamiltonian information as in the
standard configuration, supplemented with chemical
context derived from the molecular system. This
additional information includes the molecular
geometry, basis set, electronic configuration,
active space, qubit mapping, and reference state.
The circuit and Hamiltonian remain fixed, and the
model is instructed to return a complete parameter
vector without modifying either representation.

\begin{promptbox}{Physics-aware optimization - System prompt}
You optimize the expectation value of a supplied Hamiltonian
using a fixed parameterized quantum circuit.
Choose one complete absolute numerical parameter vector in radians.
Do not modify the circuit or Hamiltonian.
Infer parameter order from the OpenQASM 3 input declarations
in their textual order.
Qiskit Pauli labels run from q[n-1] on the left to q[0] on the right.
Use only the provided circuit, Hamiltonian, chemical context
(when present), optional noise model, and previous
angle-energy observations.
Return only a JSON object with keys "theta" (numeric array)
and "mechanism" (brief, testable reasoning).
\end{promptbox}

The user prompt follows the standard structure,
with an additional chemical-context section:

\begin{promptbox}{Physics-aware optimization - User prompt}
CIRCUIT_OPENQASM_3:
<parameterized OpenQASM 3 circuit>

HAMILTONIAN_PAULI_TERMS:
<Pauli strings and coefficients>

NOISE_MODEL:
<noise information, when disclosed>

PREVIOUS_EVALUATIONS:
<previous parameter vectors and evaluations, when available>

CHEMICAL_CONTEXT:
{
  "molecule": "<molecule name>",
  "geometry_angstrom": "<atomic coordinates>",
  "basis": "<basis set>",
  "charge": "<molecular charge>",
  "multiplicity": "<spin multiplicity>",
  "full_spatial_orbitals": "<number>",
  "full_num_particles": "<spin-resolved electron counts>",
  "active_spatial_orbitals": "<number>",
  "active_num_particles": "<spin-resolved active electron counts>",
  "active_orbital_indices": "<selected orbital indices>",
  "n_qubits": "<number>",
  "ansatz": "<ansatz name>",
  "mapping": "Jordan-Wigner",
  "reference_state": "Hartree-Fock; preparation included in the OpenQASM",
  "hf_basis_index": "<Hartree-Fock basis index>",
  "initial_theta": "<initial parameters, when available>",
  "shots_per_energy_query": "<number of shots>",
  "noise": "<true or false>",
  "energy_convention": "Hamiltonian Pauli coefficients describe electronic
  energy. Observed feedback uses electronic energy, excluding additive
  nuclear and frozen-core offsets."
}
\end{promptbox}

The chemical context is constructed from the
molecular problem definition rather than inferred
by the language model. Reference ground-state
energies are not included in the prompt.
The noise-model and previous-evaluation sections
are included only when applicable. Apart from
the additional chemical context, the standard
prompt structure and parameter-output format
remain unchanged.

\subsection{Circuit interpretability prompt}
\label{app:interpretability_prompt}

For blind circuit interpretation, the language model
receives an anonymized OpenQASM 3 description.
The instructions require the model to distinguish
mathematical conclusions from hypotheses about the
circuit's computational purpose and to report
uncertainty when appropriate.

\begin{promptbox}{Circuit interpretation - System prompt}
You are analyzing an unknown quantum circuit from its OpenQASM 3 description.
Use ONLY the supplied circuit. Do not use external tools, simulators, or experiments.
Do not assume a particular named algorithm. Distinguish verifiable mathematical
conclusions from hypotheses about the circuit's intended computational task.
If the algorithm's purpose is ambiguous, report the ambiguity explicitly.
Qubit indices start at q[0]; classical bit strings are written in descending
bit order, with the highest-index classical bit on the left.

Return ONLY one valid JSON object with these keys:
{
  "circuit_behavior": "Mathematical behavior of the circuit, with supporting details.",
  "algorithm_hypothesis": "Best-fitting algorithm name or 'not identifiable'.",
  "predicted_measurement_behavior": "Plain-language prediction of measurement outcomes.",
  "predicted_most_likely_bitstring": "One most likely bitstring in descending classical-bit order, or null if uncertain.",
  "evidence": ["Concrete gate-level evidence."],
  "alternative_hypotheses": ["Other possible interpretations."],
  "not_identifiable": ["Properties not identifiable from the circuit alone."],
  "confidence": 0.0
}
The example predicted_probabilities entry is illustrative, not an actual prediction.
For measured circuits, provide a normalized bitstring-to-probability object when
possible, covering all outcomes with non-negligible probability. You may provide
an empty object if you cannot infer quantitative probabilities. Never invent
an algorithm label or probabilities when uncertain. Confidence refers to the
algorithm identification, not mathematical correctness.
\end{promptbox}

The user prompt contains only the anonymized circuit:

\begin{promptbox}{Circuit interpretation -- User prompt}
OPENQASM_3:
<anonymized OpenQASM 3 circuit>
\end{promptbox}

The same instructions are used for logical, native,
and obfuscated representations. Algorithm-specific
identifiers and comments are removed before inference.
The model's predictions are subsequently compared
with independently established circuit properties.

\section{Molecular Hamiltonians and Circuit Construction}
\label{app:molecular_hamiltonians}

This appendix describes the construction of the
molecular Hamiltonians, their representation as
qubit operators, and the parameterized circuits
used in the molecular optimization experiments.

We consider neutral singlet $H_4$ and $LiH$, both described using
the STO-3G basis~\cite{Hehre1969}. For $H_4$,
we retain the full STO-3G orbital space, containing
four electrons and four spatial orbitals. For
$LiH$, we employ a reduced active space containing
two electrons and three spatial orbitals. Their atomic coordinates
are given in Appendices~\ref{app:lih_molecule}
and~\ref{app:h4_molecule}, respectively. The same Hamiltonian and circuit architecture
are used for the LMO and classical optimization
baselines within each molecular benchmark.

Within the Born-Oppenheimer approximation
\cite{born1927}, the electronic Hamiltonian of
a molecular system is expressed in second
quantization as
\begin{equation}
    H_{\mathrm{el}}
    =
    \sum_{pq} h_{pq} a_p^\dagger a_q
    +
    \frac{1}{2}
    \sum_{pqrs} h_{pqrs}
    a_p^\dagger a_q^\dagger a_s a_r,
\end{equation}
where $a_p^\dagger$ and $a_p$ are fermionic
creation and annihilation operators, respectively.
The coefficients $h_{pq}$ and $h_{pqrs}$ are
the one- and two-electron integrals in the
chosen orbital basis. The molecular energy
additionally includes nuclear repulsion and,
when an active-space approximation is employed,
the energy associated with inactive electrons.

The electronic Hamiltonian is mapped to qubit
operators using the Jordan-Wigner
transformation~\cite{jordan_uber_1928}.
The resulting qubit Hamiltonian takes the form
\begin{equation}
    H_{\mathrm{qubit}}
    =
    \sum_{j=1}^{M} c_j P_j,
\end{equation}
where $c_j$ are real coefficients and
$P_j$ are tensor products of Pauli operators.
The physical molecular energy is calculated as
\begin{equation}
    E_{\mathrm{mol}}(\boldsymbol{\theta})
    =
    \langle \psi(\boldsymbol{\theta})|
    H_{\mathrm{qubit}}
    |\psi(\boldsymbol{\theta})\rangle
    +
    E_{\mathrm{offset}},
\end{equation}
where $E_{\mathrm{offset}}$ accounts for
constant energy contributions excluded
from the mapped qubit Hamiltonian.

The variational state is prepared using a fixed
parameterized quantum circuit,
\begin{equation}
    |\psi(\boldsymbol{\theta})\rangle
    =
    U(\boldsymbol{\theta})
    |\psi_{\mathrm{ref}}\rangle,
\end{equation}
where $|\psi_{\mathrm{ref}}\rangle$ denotes
the Hartree--Fock reference state and
$\boldsymbol{\theta}$ contains the variational
parameters. For the $LiH$ benchmark, we employ
a unitary coupled-cluster ansatz with single
and double excitations
(UCCSD)~\cite{Romero_2019}.
The circuit architecture remains unchanged
during optimization, with both the LMO and
classical optimizers modifying only the
numerical parameter values.

For LMO inference, the parameterized circuit
is serialized into OpenQASM 3, and the
Hamiltonian is provided as a list of Pauli
strings and their coefficients. The parameter
ordering is explicitly defined to ensure
consistency between the generated values
and the circuit. Optimization performance
is evaluated using the energy error relative
to the corresponding full configuration
interaction (FCI) reference,
\begin{equation}
    \Delta E(\boldsymbol{\theta})
    =
    E_{\mathrm{mol}}(\boldsymbol{\theta})
    -
    E_{\mathrm{FCI}}.
\end{equation}
The reference energy is obtained by
diagonalizing the molecular Hamiltonian
within the relevant physical electron-number
and spin sector. We adopt the conventional
chemical-accuracy threshold of $1$ kcal/mol,
corresponding approximately to
$\epsilon_{\mathrm{chem}}
\simeq 1.59\times10^{-3}\,E_h$.

\section{$LiH$ Molecule}
\label{app:lih_molecule}

We further evaluate the LMO on the molecular
ground-state energy problem of lithium hydride
($LiH$) molecule, which is defined by the
following Cartesian coordinates in angstroms:
\begin{equation}
\mathrm{Li}:\ (0,0,0),\qquad
\mathrm{H}:\ (0,0,1.60).
\label{eq:lih_geometry}
\end{equation}
The electronic Hamiltonian is constructed
using the STO-3G basis \cite{Hehre1969} and an active-space
approximation containing two electrons. The
fermionic Hamiltonian is mapped to qubit operators
using the Jordan-Wigner transformation \cite{jordan_uber_1928}. The
variational state is prepared using a
unitary coupled-cluster ansatz with single and
double excitations (UCCSD) \cite{Romero_2019}, initialized from the
Hartree-Fock reference state. The circuit
architecture remains fixed throughout optimization
with only its numerical parameters being updated.

\begin{figure}[H]
    \centering
    \includegraphics[width=1.0\textwidth]
    {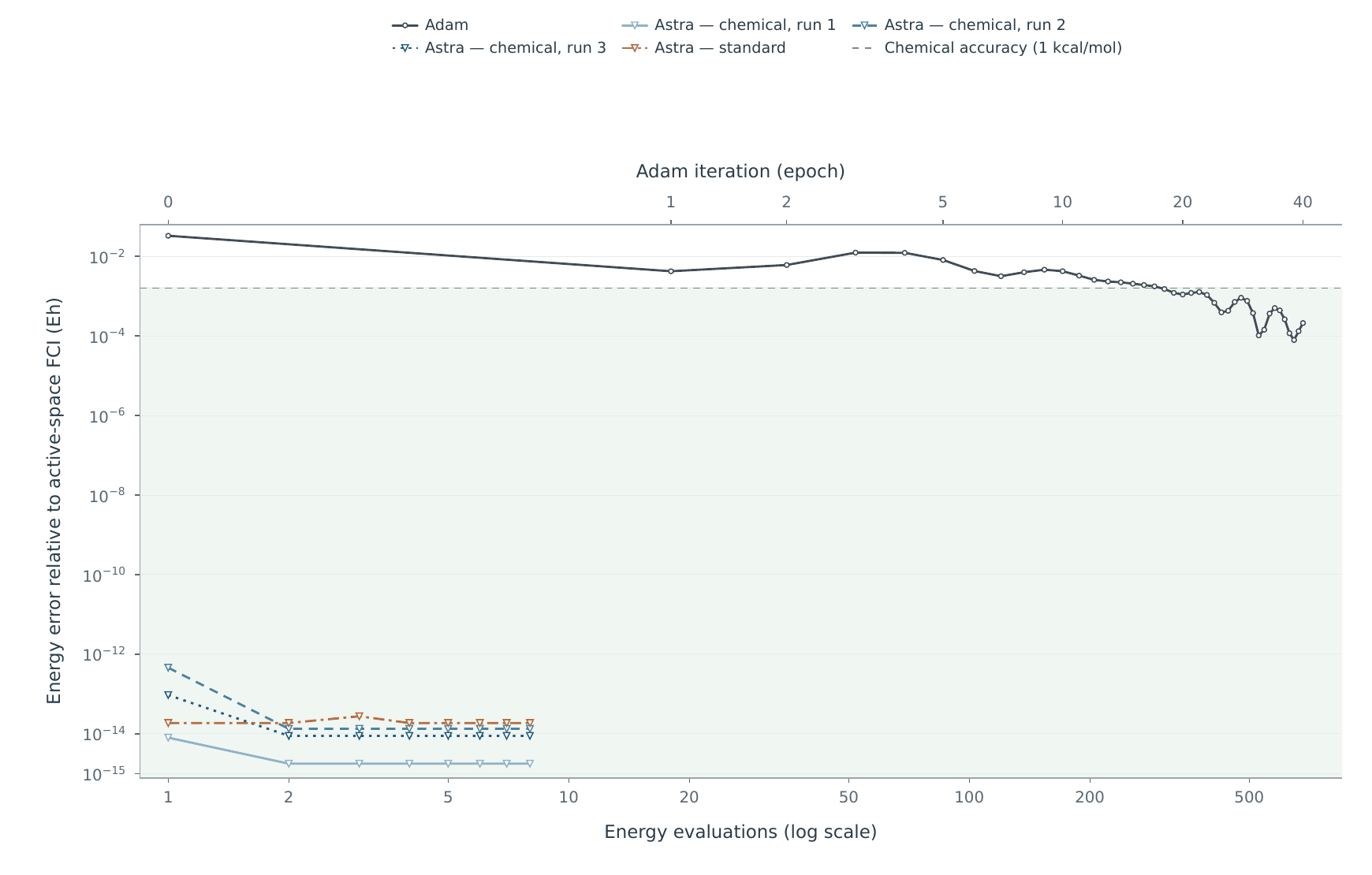}
    \caption{\justifying
    Optimization of the $LiH$ molecular Hamiltonian
    using Adam and the language model optimizer. Three runs with the physics-aware
    (chemical) prompt and one run with the
    standard blind prompt are shown.
    The vertical axis represents the energy
    error relative to the active-space full
    configuration interaction reference,
    while the horizontal axis shows the
    cumulative number of Hamiltonian energy
    evaluations on a logarithmic scale.
    The dashed horizontal line indicates the
    chemical-accuracy threshold of $1$ kcal/mol.
    The upper horizontal axis indicates the
    corresponding Adam optimization iterations.
    }
    \label{fig:lih_gap}
\end{figure}

As in the $H_4$ experiment, we investigate the
standard blind and physics-aware prompting
configurations for the lithium hydride
($LiH$) molecule. Both use the same frozen LLM
to generate complete parameter vectors, which
are evaluated using the molecular Hamiltonian.
The physics-aware configuration is evaluated
in three repeated runs, while one standard
blind-prompt run is included for comparison.
The resulting trajectories are compared with
Adam using the cumulative number of Hamiltonian
energy evaluations. Figure~\ref{fig:lih_gap} presents the energy errors relative to the active-space full
configuration interaction (FCI) reference.
All four LMO runs achieve energy errors near
numerical precision within the first few
evaluations. In contrast, Adam begins with
a substantially larger energy error and
requires considerably more objective evaluations
to reach the chemical-accuracy threshold.

The LMO trajectories remain close to the
reference energy across subsequent proposals,
with only small differences between repeated
runs. These results demonstrate that the
LMO can generate near-ground-state parameter
configurations for the tested $LiH$ problem
with a small number of objective evaluations.
However, the comparison concerns quantum
objective-evaluation requirements rather than
total computational cost, and the limited
number of runs does not establish a systematic
advantage of physics-aware prompting over
the blind configuration.

\section{$H_4$ Molecule}
\label{app:h4_molecule}

We examine two geometries of neutral singlet $H_4$
using STO-3G. The standard reference structure is a
planar rectangle, while the second is a nonplanar
and in-plane distorted structure. Both configurations
contain four electrons in four spatial orbitals,
without frozen-core or active-space truncation. Their
coordinates are given in Table~\ref{tab:h4_geometries}.

\begin{table}[htb]
\caption{Cartesian geometries of the standard and
distorted $H_4$ molecules, in angstroms.}
\label{tab:h4_geometries}
\begin{tabular}{crrr@{\qquad}rrr}
\toprule
& \multicolumn{3}{c}{Standard $H_4$} &
  \multicolumn{3}{c}{Distorted $H_4$}\\
  \cmidrule(lr){2-4} \cmidrule(lr){5-7}
Atom & $x$ & $y$ & $z$ & $x$ & $y$ & $z$\\
\hline
H$_1$ & 0.00 & 0.00 & 0.00 & 0.00 & 0.00 & 0.00\\
H$_2$ & 1.06 & 0.00 & 0.00 & 1.00 & 0.00 & 0.00\\
H$_3$ & 0.00 & 1.20 & 0.00 & 0.00 & 1.20 & 0.20\\
H$_4$ & 1.06 & 1.20 & 0.00 & 1.05 & 1.20 & $-0.20$\\
\bottomrule
\end{tabular}
\end{table}

The standard geometry is the one in the main text,
Fig.~\ref{fig:h4_gap}. The distorted geometry was
introduced as an additional generalization probe.
In particular, the blind-prompt test assesses whether
parameter proposals that may exploit familiar molecular
patterns remain effective if
Hamiltonian differs from that of the standard
configuration.
This comparison is intended to test sensitivity to
the molecular instance.
Because the chemical-context prompt discloses the
molecular geometry and electronic setting, the blind
configuration is especially relevant when examining
the response to the modified Hamiltonian
without that information.

\begin{figure}[htbp]
    \centering
    \includegraphics[width=\linewidth]
    {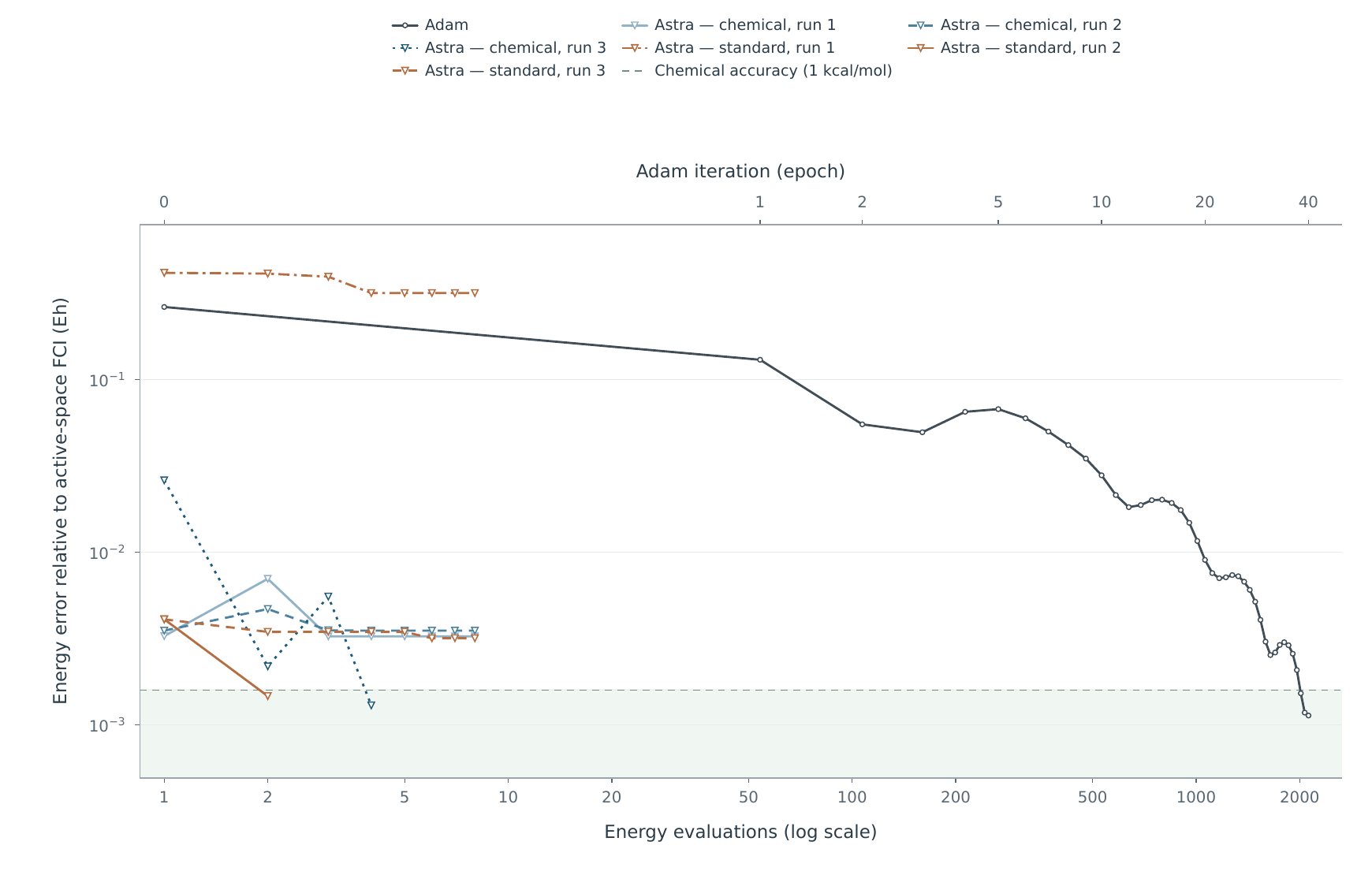}
    \caption{\justifying
    Optimization trajectories for the distorted
    $H_4$ Hamiltonian corresponding to the nonplanar
    coordinates in Table~\ref{tab:h4_geometries}.
    The fixed eight-qubit, 26-parameter UCCSD circuit
    is optimized with the standard (blind) and
    physics-aware (chemical) LMO prompts, with three
    independent runs per prompt, and compared with Adam.
    The vertical axis shows the energy error from the
    FCI reference of the distorted Hamiltonian; the
    lower horizontal axis gives cumulative Hamiltonian
    energy evaluations on a logarithmic scale. The
    upper axis shows Adam iterations. The dashed line
    and shaded region mark chemical accuracy
    $1\,\mathrm{kcal\,mol^{-1}}$.}
    \label{fig:h4_distorted_gap}
\end{figure}

The results for the planar geometry appear in
Fig.~\ref{fig:h4_gap} and the distorted-geometry
trajectories appear in Fig.~\ref{fig:h4_distorted_gap}.
In the distorted case, the quality of LMO-generated
parameters varies substantially across independent
runs. Some runs approach or cross the chemical-accuracy
threshold, while one blind-prompt trajectory remains
well above it. These results
illustrate the sensitivity of the LMO's
performance to the molecular Hamiltonian,
although they do not, by themselves, establish
whether the model relies on previously seen data.

\section{Random Parent Hamiltonian}

We construct a family of variational
optimization problems using parent Hamiltonians,
following the benchmark construction principle
of Kobayashi et al.~\cite{kobayashi2022parent}.
This approach provides Hamiltonians with known
ground-state energies whose ground states
are guaranteed to be representable by the
chosen parameterized circuits. The ansatz $U(\boldsymbol{\theta})$
consists of alternating blocks of single-qubit $R_Y R_Z R_Y$
rotations and fixed nearest-neighbor $R_{XX}(\pi/2)$ entangling
gates. For each instance, a hidden parameter vector
$\boldsymbol{\theta}^{*}$ is sampled and used to construct
\begin{equation}
    H =
    U(\boldsymbol{\theta}^{*})
    H_0
    U^{\dagger}(\boldsymbol{\theta}^{*}),
    \qquad
    H_0 = \sum_{q=0}^{n-1}\frac{I-Z_q}{2}.
    \label{eq:random_parent_hamiltonian}
\end{equation}

Since unitary conjugation preserves the spectrum, $H$ has a
unique ground state
$U(\boldsymbol{\theta}^{*})\ket{0}^{\otimes n}$,
ground-state energy $E_0=0$, and spectral gap equal to one. To control the size of the Pauli decomposition, the hidden
parameters are sampled predominantly from the Clifford angles
$\{0,\pm\pi/2,\pi\}$, while $r$ selected rotations receive
non-Clifford angles. This limits the proliferation of Pauli
terms while retaining a continuously parameterized optimization
circuit. Instances are accepted subject to limits on the number
of nonzero Pauli terms and compiled circuit depth. The LMO
receives the parameterized circuit and the resulting Pauli
Hamiltonian, but neither the hidden parameters nor the
construction procedure. Its objective is to find any parameter
vector that prepares a low-energy state, rather than to recover
the original hidden angles. Further construction details are
provided in Appendix~\ref{app:parent_hamiltonian}.

We evaluate the LMO on circuits with $n\in\{4,8,12\}$ qubits,
$L\in\{1,2,4\}$ entangling layers, and
$r\in\{1,2\}$ non-Clifford rotations in the generating
circuit. Five independently generated instances are considered
for each configuration, yielding 90 optimization problems.
For each instance, the frozen LMO proposes a complete parameter
vector without receiving any objective-function evaluations.
Subsequent proposals use feedback from previous
energy evaluations, with up to five proposals per instance.
All energies in this experiment are evaluated using exact,
noiseless statevector simulations. We define a successful
solution by an energy gap $E(\boldsymbol{\theta})-E_0
\leq 10^{-6}$.

\begin{figure*}[h]
    \centering
    \begin{subfigure}{0.49\textwidth}
        \centering
        \includegraphics[width=\linewidth]
        {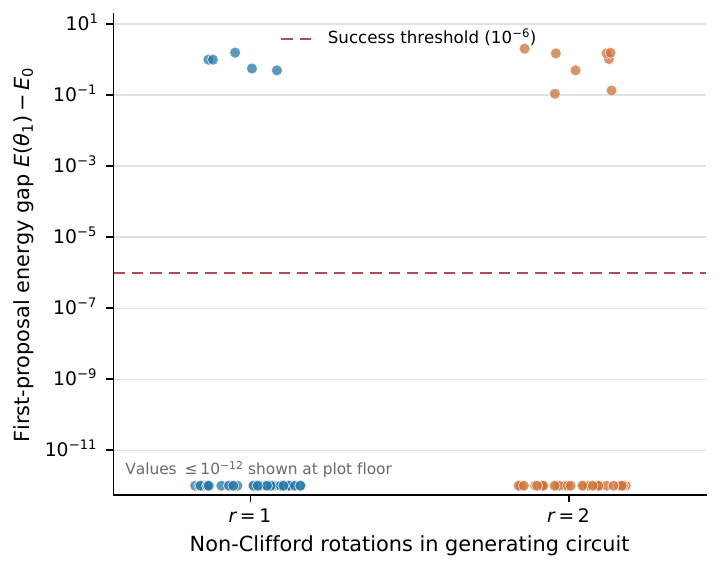}
        \caption{}
        \label{fig:parent_hamiltonian_a}
    \end{subfigure}
    \hfill
    \begin{subfigure}{0.49\textwidth}
        \centering
        \includegraphics[width=\linewidth]
        {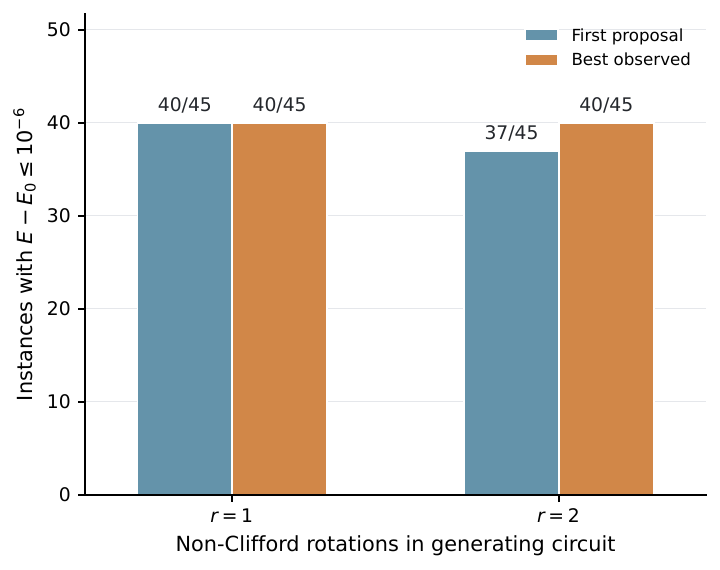}
        \caption{}
        \label{fig:parent_hamiltonian_b}
    \end{subfigure}
    \caption{
        \justifying Performance of the language model optimizer on
        random parent Hamiltonians with $r=1$ and $r=2$
        non-Clifford rotations in the generating circuit.
        (a) Distribution of first-proposal energy gaps
        for 45 instances per family, obtained without
        prior objective-function feedback. The dashed
        line indicates the success threshold of $10^{-6}$;
        gaps below $10^{-12}$ are displayed at the
        numerical plotting floor.
        (b) Number of instances reaching the success
        threshold on the first proposal and at any
        observed optimization round.
        All energies are evaluated using exact,
        noiseless simulations.
    }
    \label{fig:parent_hamiltonian_main}
\end{figure*}

Figure~\ref{fig:parent_hamiltonian_main}(a) shows the
distribution of first-proposal energy gaps across the 90
instances. Without prior objective feedback, the LMO reaches
the target accuracy in 40 of 45 instances ($88.9\%$) for
$r=1$ and 37 of 45 instances ($82.2\%$) for $r=2$.
Most successful proposals achieve energy gaps near numerical
precision, whereas the unsuccessful instances exhibit
substantially larger errors. These results demonstrate that
the LMO can generate near-ground-state parameters directly
from symbolic circuit and Hamiltonian descriptions for a
substantial fraction of the tested problems.

Allowing the LMO to incorporate energy feedback increases
the number of successful instances for $r=2$ from 37 to 40,
as shown in Fig.~\ref{fig:parent_hamiltonian_main}(b).
For $r=1$, the number of successful instances remains at
40, although feedback reduces the energy in some initially
unsuccessful cases. Across both families, the best observed
proposals reach the target accuracy in 80 of 90 instances
($88.9\%$). These observations indicate that feedback can
refine initially unsuccessful proposals, although its effect
varies across instances.

A comparison with the gradient-based Adam optimizer
further highlights the difference in objective-evaluation
requirements. While Adam generally requires repeated
parameter-shift gradient evaluations to approach low-energy
solutions, the LMO frequently identifies near-ground-state
parameters in its first proposal, without prior objective
feedback. This comparison concerns the number of
Hamiltonian evaluations rather than total computational
cost, as the LMO additionally performs language-model
inference and has access to the symbolic problem description.
Detailed optimization trajectories and resource accounting
are presented in Appendix~\ref{app:parent_classical}.

Further structural analysis shows that successful
state preparation does not generally require recovery
of the hidden generating parameters. The LMO often
achieves near-unit ground-state fidelity despite
substantial differences between the proposed and
planted parameter vectors. Moreover, rounding the
proposed angles to Clifford-compatible values causes
all 90 first proposals to exceed the success threshold,
indicating the importance of continuous rotation values.
The model-generated explanations also contain
physically testable claims, although their consistency
with numerical calculations is not guaranteed.
These findings are specific to the structured
parent-Hamiltonian family and do not establish a
general computational advantage. Additional analyses
are provided in Appendix~\ref{app:parent_structural}.

\section{Parent Hamiltonian Construction}
\label{app:parent_hamiltonian}

This appendix describes the parameterized circuit, the
generation of random parent Hamiltonians, and the constraints
used to obtain instances with a controlled number of Pauli
terms. The construction guarantees an exactly representable
ground state while allowing the complexity of the generated
Hamiltonian to vary independently of the number of trainable
circuit parameters.

\subsection{Parameterized Circuit}

We consider an $n$-qubit circuit $U(\boldsymbol{\theta})$ acting
on the initial state $\ket{0}^{\otimes n}$, where
$\boldsymbol{\theta}\in\mathbb{R}^{p}$ contains the trainable
rotation angles. The resulting state is
\begin{equation}
    \ket{\psi(\boldsymbol{\theta})}
    =
    U(\boldsymbol{\theta})\ket{0}^{\otimes n}.
\end{equation}

The ansatz alternates complete single-qubit rotation blocks
with fixed entangling layers. Each rotation block applies
$R_Y R_Z R_Y$ to every qubit. In chronological gate order,
the corresponding unitary is
\begin{equation}
    B_{\ell}(\boldsymbol{\theta}_{\ell})
    =
    \bigotimes_{q=0}^{n-1}
    R_Y(\gamma_{\ell,q})
    R_Z(\beta_{\ell,q})
    R_Y(\alpha_{\ell,q}),
    \label{eq:parent_rotation_block}
\end{equation}
where $\alpha_{\ell,q}$, $\beta_{\ell,q}$, and
$\gamma_{\ell,q}$ are independent parameters. The rightmost
operator acts first.

The entangling layers contain fixed nearest-neighbor gates,
\begin{equation}
    E_{\ell}
    =
    \prod_{\substack{
        q=\ell\bmod 2\\
        q<n-1
    }}^{\mathrm{step}\;2}
    R_{XX}^{(q,q+1)}
    \left(\frac{\pi}{2}\right),
\end{equation}
where
\begin{equation}
    R_{XX}(\phi)
    =
    \exp\left(
        -i\frac{\phi}{2}X\otimes X
    \right).
\end{equation}

The interacting pairs alternate between even- and odd-indexed
connections. For $L$ entangling layers, the complete circuit
is
\begin{equation}
    U(\boldsymbol{\theta})
    =
    B_L E_{L-1} B_{L-1}
    \cdots E_1 B_1 E_0 B_0.
    \label{eq:parent_ansatz}
\end{equation}

There are $L+1$ rotation blocks, giving $p=3n(L+1)$
trainable parameters. All rotation angles remain independently
adjustable during optimization.

\subsection{Reference and Parent Hamiltonians}

The reference Hamiltonian is defined as
\begin{equation}
    H_0
    =
    \sum_{q=0}^{n-1}\frac{I-Z_q}{2},
    \label{eq:parent_reference}
\end{equation}
where $Z_q$ acts on qubit $q$. Its eigenstates are the computational-basis states, with
eigenvalues equal to their Hamming weights:
\begin{equation}
    H_0\ket{x}
    =
    w_H(x)\ket{x},
    \qquad x\in\{0,1\}^{n}.
\end{equation}

Consequently, $H_0$ has the unique ground state
$\ket{0}^{\otimes n}$, ground-state energy zero, and spectral
gap one. For each benchmark instance, a hidden parameter vector
$\boldsymbol{\theta}^{*}$ is selected and inserted into the
ansatz. The resulting unitary
$U^{*}=U(\boldsymbol{\theta}^{*})$ defines the parent
Hamiltonian
\begin{equation}
    H
    =
    U^{*}H_0U^{*\dagger}.
    \label{eq:parent_construction}
\end{equation}

By construction,
\begin{equation}
    \begin{aligned}
    H U^{*}\ket{0}^{\otimes n}
    &=
    U^{*}H_0U^{*\dagger}
    U^{*}\ket{0}^{\otimes n}\\
    &=
    U^{*}H_0\ket{0}^{\otimes n}\\
    &=0.
    \end{aligned}
\end{equation}

Therefore,
\begin{equation}
    \ket{\psi^{*}}
    =
    U^{*}\ket{0}^{\otimes n}
\end{equation}
is the unique ground state of $H$. Because unitary conjugation preserves the spectrum, the exact
parent Hamiltonian retains the eigenvalues of $H_0$, including
the zero ground-state energy and unit spectral gap.

\subsection{Sparse Non-Clifford Parameter Sampling}

Sampling every hidden rotation angle continuously can produce
a large Pauli expansion after conjugation. To control this
growth, we use a sampling procedure that combines Clifford
rotations with a configurable number of non-Clifford rotations. Initially, each hidden angle is sampled from
\begin{equation}
    \mathcal{C}
    =
    \left\{
        0,\frac{\pi}{2},
        -\frac{\pi}{2},\pi
    \right\}.
    \label{eq:clifford_angle_set}
\end{equation}

A subset
$\mathcal{S}\subseteq\{1,\ldots,p\}$ of size
$|\mathcal{S}|=r$ is then selected. The corresponding angles
are replaced by values sampled from $[-\pi,\pi)$, excluding
Clifford-valued rotations. Thus, the hidden parameters follow
\begin{equation}
    \theta_j^{*}
    \in
    \begin{cases}
        [-\pi,\pi)\setminus\mathcal{C},
            & j\in\mathcal{S},\\
        \mathcal{C},
            & j\notin\mathcal{S},
    \end{cases}
    \label{eq:sparse_hidden_sampling}
\end{equation}
with angles understood modulo $2\pi$. The parameter $r$ controls the number of non-Clifford
rotations in the hidden generating circuit. It does not
restrict the optimization domain: all $p$ parameters remain
continuous and trainable. This distinction follows from the action of Clifford
operations on Pauli operators. For any Clifford unitary $C$
and Pauli string $P$,
\begin{equation}
    CPC^{\dagger}=\pm P',
\end{equation}
where $P'$ is another Pauli string. Consequently, Clifford conjugation does not increase the
number of Pauli strings. In particular, when $r=0$, the
reference Hamiltonian is mapped to a sum of $n$ nonidentity
Pauli terms and one identity term. A non-Clifford Pauli rotation, by contrast, can split an
anticommuting Pauli operator into two components. For
example,
\begin{equation}
    R_Y(\theta) Z R_Y^{\dagger}(\theta)
    =
    \cos(\theta)Z+\sin(\theta)X.
\end{equation}

Each non-Clifford rotation can therefore increase the term
count by at most a factor of two. Before collecting repeated
Pauli strings, the construction admits the bound
\begin{equation}
    M\leq 1+n2^r,
    \label{eq:parent_term_bound}
\end{equation}
where $M$ includes the identity term. The actual number of nonzero terms may be considerably
smaller because some Pauli strings commute with the applied
rotations and repeated contributions can combine or cancel. The generated instances are also checked to avoid counting
selected non-Clifford angles that have no effect on the
resulting Hamiltonian. Thus, the sampling parameter $r$
describes the intended non-Clifford content of accepted
instances, rather than merely the number of continuous
angles sampled.

\subsection{Pauli Decomposition}

The parent Hamiltonian is represented as
\begin{equation}
    H
    =
    \sum_{j=1}^{M}c_jP_j,
    \qquad
    P_j\in\{I,X,Y,Z\}^{\otimes n},
    \label{eq:parent_pauli_representation}
\end{equation}
with real coefficients $c_j$.

To obtain this representation, the reference Pauli terms are
conjugated sequentially through the hidden circuit. For a
gate of the form
\begin{equation}
    R_P(\phi)
    =
    \exp\left(-i\frac{\phi}{2}P\right),
\end{equation}
where $P$ is a Hermitian Pauli string, the conjugation of
another Pauli string $Q$ satisfies
\begin{equation}
    R_P(\phi) Q R_P^{\dagger}(\phi)
    =
    \begin{cases}
        Q, & [P,Q]=0,\\[2mm]
        Q\cos\phi-iPQ\sin\phi,
            & \{P,Q\}=0.
    \end{cases}
    \label{eq:pauli_conjugation_rule}
\end{equation}

This identity permits the Hamiltonian to be constructed using
Pauli-string algebra, without explicitly forming a dense
$2^n\times2^n$ matrix. After combining identical strings, coefficients below a
specified numerical tolerance are discarded. The resulting
term count is evaluated against the prescribed budget $K$:
\begin{equation}
    M\leq K.
\end{equation}

Instances are additionally subject to a maximum compiled
native-gate depth $D_{\max}$:
\begin{equation}
    D_{\mathrm{native}}\leq D_{\max}.
\end{equation}

Candidate instances that do not satisfy the generation
constraints are rejected, and new hidden parameters are
sampled up to a specified attempt limit. These constraints control the size of the Hamiltonian
representation and the circuit resources. They do not, by
themselves, establish the computational difficulty of an
accepted optimization problem.

\subsection{Numerical Benchmark Configuration}
\label{app:parent_benchmark_configuration}

The numerical benchmark spans three system sizes,
$n\in\{4,8,12\}$, three entangling-layer counts,
$L\in\{1,2,4\}$, and two non-Clifford rotation counts,
$r\in\{1,2\}$. For each combination, five instances
are generated using seeds $42$--$46$, resulting in
$3\times3\times2\times5=90$ optimization problems.
Table~\ref{tab:parent_benchmark_configuration} summarizes
these settings and the constraints imposed during generation.

\begin{table}[t]
    \centering
    \caption{Configuration of the random parent-Hamiltonian
    benchmark. The Pauli-term budget includes the identity
    contribution.}
    \label{tab:parent_benchmark_configuration}
    \begin{tabular}{ll}
        \hline\hline
        Setting & Value \\
        \hline
        Qubits $n$ & $\{4,8,12\}$ \\
        Entangling layers $L$ & $\{1,2,4\}$ \\
        Non-Clifford rotations $r$ & $\{1,2\}$ \\
        Random seeds & $42$-$46$ \\
        Trainable parameters $p$ & $3n(L+1)$ \\
        Pauli-term budget $K$ & $16$ ($r=1$); $32$ ($r=2$) \\
        Maximum compiled depth $D_{\max}$ & $100$ \\
        Maximum generation attempts & $250$ \\
        Number of instances & $90$ \\
        \hline\hline
    \end{tabular}
\end{table}

For each candidate, the hidden angles are sampled according
to Eq.~\eqref{eq:sparse_hidden_sampling}, the Pauli
coefficients are constructed using
Eq.~\eqref{eq:pauli_conjugation_rule}, and the resulting
instance is checked against the term-count and compiled-depth
limits. Candidates failing these constraints are rejected.
The numerical coefficient cutoff and the native compilation
basis are implementation-dependent settings and should be
reported alongside the released generation code.

The resulting instances have known reference energies and
exactly representable ground states, but they retain the
structure of the circuit family used to generate them. They
should not be interpreted as representative of arbitrary
physical Hamiltonians. Moreover, neither the Pauli-term count
nor the number of hidden non-Clifford rotations is, by itself,
a measure of optimization difficulty. Classical optimization
comparisons and additional structural analyses are reported
in the following appendices.

\section{Classical Optimization Comparisons}
\label{app:parent_classical}

This appendix compares the language model optimizer (LMO)
with classical optimization methods on the parent-Hamiltonian
benchmarks defined in Appendix~\ref{app:parent_hamiltonian}.
We compare the LMO with Adam as a conventional
gradient-based reference, examining solution quality
and the number of full-Hamiltonian expectation-value
evaluations required.

\subsection{Optimization Protocols and Resource Accounting}

For each benchmark instance, the optimization objective is
\begin{equation}
    E(\boldsymbol{\theta})
    =
    \bra{\psi(\boldsymbol{\theta})}
    H
    \ket{\psi(\boldsymbol{\theta})},
\end{equation}
where the exact ground-state energy is $E_0=0$.
All comparisons use exact, noiseless statevector
simulations. A solution is considered successful when
$E(\boldsymbol{\theta})-E_0\leq 10^{-6}$.

The LMO receives the parameterized circuit and its Pauli
Hamiltonian as symbolic inputs. Its first parameter proposal
is generated without previous objective evaluations.
Subsequent proposals may incorporate the history of evaluated
parameters and their energies. Each proposal specifies the
complete parameter vector and is evaluated using the same
Hamiltonian objective.

For comparison, we employ Adam with gradients computed
through the parameter-shift rule. For a trainable parameter
$\theta_j$ associated with a Pauli rotation,
\begin{equation}
    \frac{\partial E}{\partial\theta_j}
    =
    \frac{1}{2}
    \left[
        E\left(\boldsymbol{\theta}
        +\frac{\pi}{2}\boldsymbol{e}_j\right)
        -
        E\left(\boldsymbol{\theta}
        -\frac{\pi}{2}\boldsymbol{e}_j\right)
    \right],
    \label{eq:parent_parameter_shift}
\end{equation}
where $\boldsymbol{e}_j$ denotes the corresponding coordinate
vector. The ansatz contains $p=3n(L+1)$ trainable
parameters, so a direct gradient calculation requires
$2p$ shifted full-Hamiltonian expectation evaluations.
Additional evaluations of the unshifted objective are
counted separately.

We report cumulative full-Hamiltonian energy evaluations
rather than optimizer iterations alone. Each LMO proposal
requires one evaluation to assess its energy, whereas
each Adam update requires multiple evaluations to compute
the gradient. The computational cost of language-model
inference and other classical computations is not represented
by this evaluation count.

The methods also use different information. The LMO
receives the explicit circuit and Hamiltonian descriptions,
whereas Adam accesses the objective through numerical
energy and gradient evaluations. Accordingly, differences
in objective-evaluation counts should not be interpreted
as differences in total computational cost or as evidence
of a general computational advantage.

Adam is evaluated for up to 200 optimization steps, the configurations are summarized in
Table~\ref{tab:parent_optimizers}.

\begin{table}[t]
    \centering
    \caption{
        Optimization configurations used for the
        parent-Hamiltonian comparisons.
        All experiments use exact, noiseless energies.
    }
    \label{tab:parent_optimizers}
    \begin{tabular}{lcc}
        \hline\hline
        Configuration & LMO & Adam \\
        \hline
        Parameter initialization
            & Model-generated & Seed 123 \\
        Initialization scale
            & -- & 0.3 \\
        Optimization rounds
            & Up to 5 & 200 \\
        Learning rate, $r=1$
            & -- & 0.1 \\
        Learning rate, $r=2$
            & -- & 0.01 \\
        Gradient method
            & None & Parameter shift \\
        Shots per energy query
            & 0 & 0 \\
        Noise
            & None & None \\
        \hline\hline
    \end{tabular}
\end{table}

\subsection{Comparison with Adam}

\begin{figure*}[h]
    \centering
    \includegraphics[width=\textwidth]
    {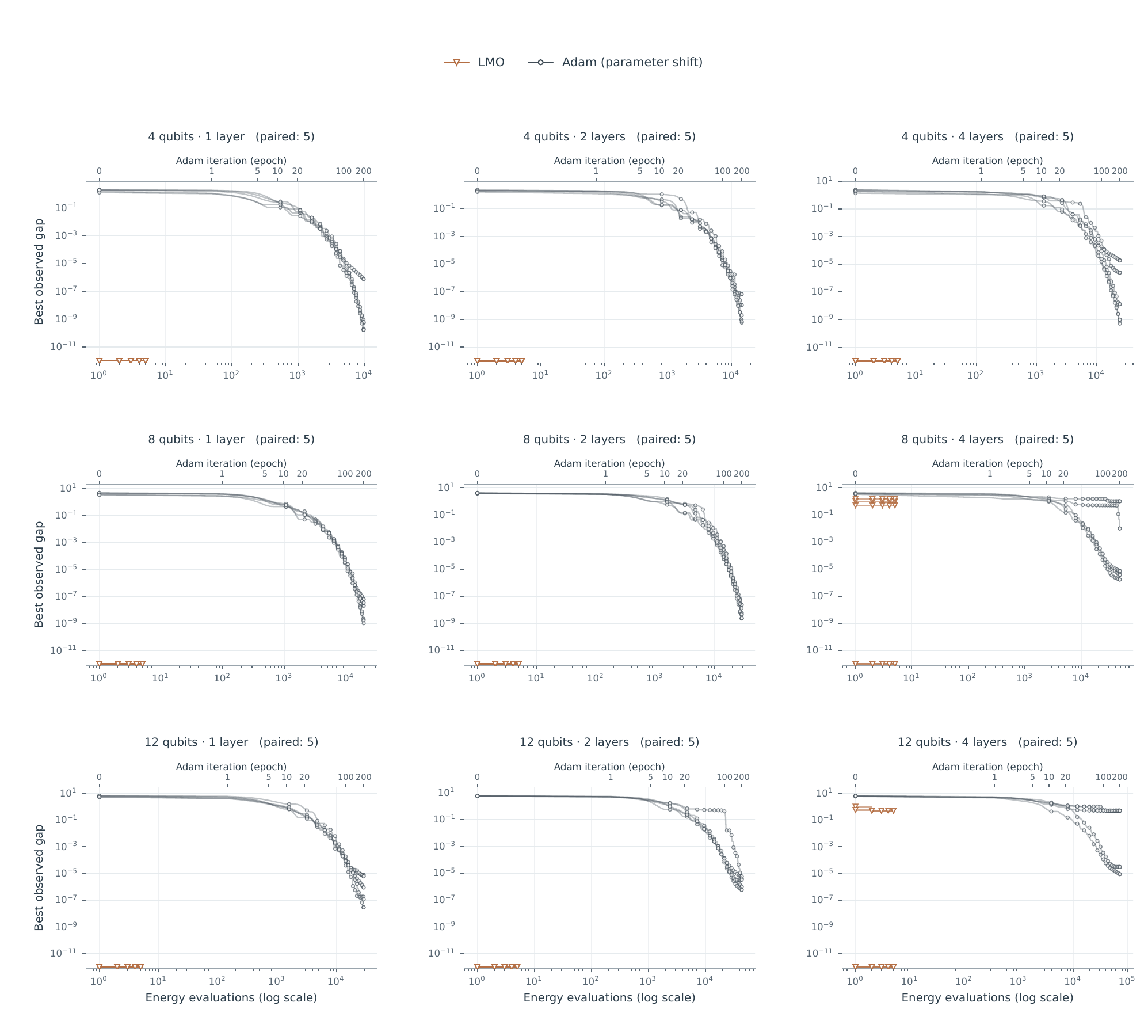}
    \caption{ \justifying
        Optimization trajectories for parent Hamiltonians
        with $r=1$ non-Clifford rotation in the generating
        circuit. Each panel corresponds to a configuration
        of qubit count and entangling layers and includes
        the available paired instances. The curves show
        the best-observed energy gaps against cumulative
        full-Hamiltonian evaluations. The upper axis
        indicates Adam iterations.
    }
    \label{fig:parent_trajectories_r1}
\end{figure*}

\begin{figure*}[h]
    \centering
    \includegraphics[width=\textwidth]
    {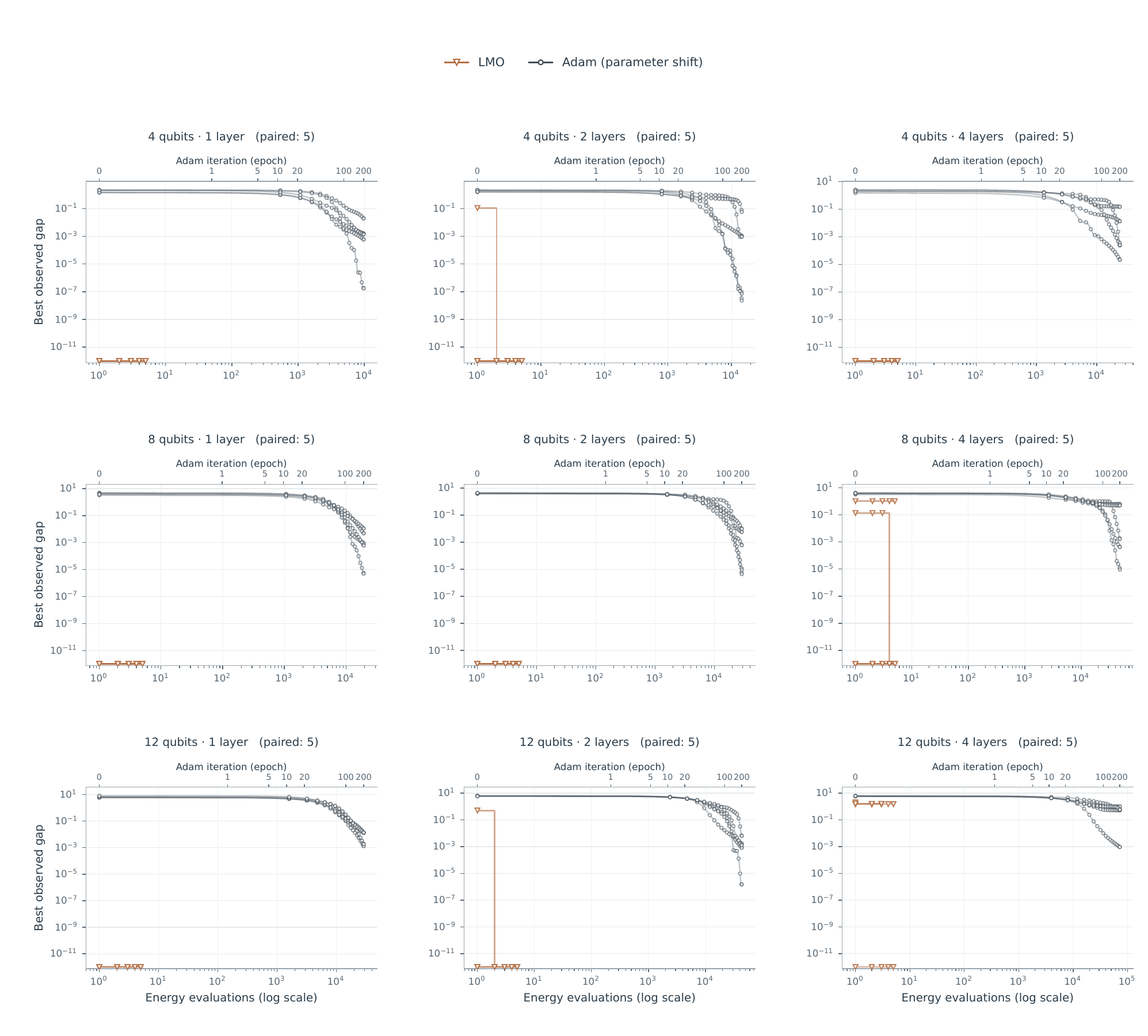}
    \caption{ \justifying
        Optimization trajectories for parent Hamiltonians
        with $r=2$ non-Clifford rotations in the generating
        circuit. The plotting conventions are the same as
        in Fig.~\ref{fig:parent_trajectories_r1}.
    }
    \label{fig:parent_trajectories_r2}
\end{figure*}

We compare LMO and Adam across configurations with
$n\in\{4,8,12\}$ qubits, $L\in\{1,2,4\}$ entangling
layers, and $r\in\{1,2\}$ non-Clifford rotations in the
generating circuit. Figures~\ref{fig:parent_trajectories_r1}
and~\ref{fig:parent_trajectories_r2} present the optimization
trajectories for the two benchmark families.

The vertical axis reports the best energy gap observed
up to each recorded point,
\begin{equation}
    \Delta E_{\mathrm{best}}(t)
    =
    \min_{s\leq t}
    \left[
        E(\boldsymbol{\theta}_s)-E_0
    \right].
    \label{eq:parent_best_gap}
\end{equation}
The lower horizontal axis gives the cumulative number of
full-Hamiltonian energy evaluations, while the upper axis
indicates the corresponding Adam optimization iterations.
For LMO, each recorded point corresponds to a complete
parameter proposal evaluated using the Hamiltonian
objective.

We can observe that in several configurations, Adam continues reducing the energy gap well beyond the initial iterations,
eventually approaching the ground-state energy. In contrast, the LMO frequently reaches near-ground-state
energies in its first proposal, without previous objective
feedback. For these instances, the energy gaps are close
to numerical precision after a single Hamiltonian
evaluation. However, the LMO does not succeed universally,
and some configurations exhibit substantially larger
residual errors.

The comparison highlights the difference in
Hamiltonian-evaluation requirements between the tested
methods. Adam can obtain low-energy solutions through
continued gradient-based optimization, whereas the LMO
often produces suitable parameters before receiving
objective feedback. These observations concern the
specific structured benchmarks and optimizer
configurations considered here, rather than a general
comparison of computational efficiency. The parent Hamiltonian instances possess a planted structure that may be exploitable by specialized classical algorithms. The comparison with Adam therefore characterizes objective-evaluation requirements relative to a conventional gradient-based optimizer, rather than establishing classical hardness or a general computational advantage.

\section{Structural Analysis of LMO Parameter Proposals}
\label{app:parent_structural}

This appendix examines the structure of the parameter
vectors generated by the language model optimizer (LMO)
and the physical explanations accompanying its proposals.
We investigate the role of non-Clifford rotations,
the relationship between the proposed and generating
parameter vectors, and the physical consistency of
selected model-generated explanations.

\subsection{Clifford-Angle Ablation and Parameter Non-Uniqueness}

\begin{figure*}[t]
    \centering
    \begin{subfigure}[t]{0.49\textwidth}
        \centering
        \includegraphics[width=\linewidth]
        {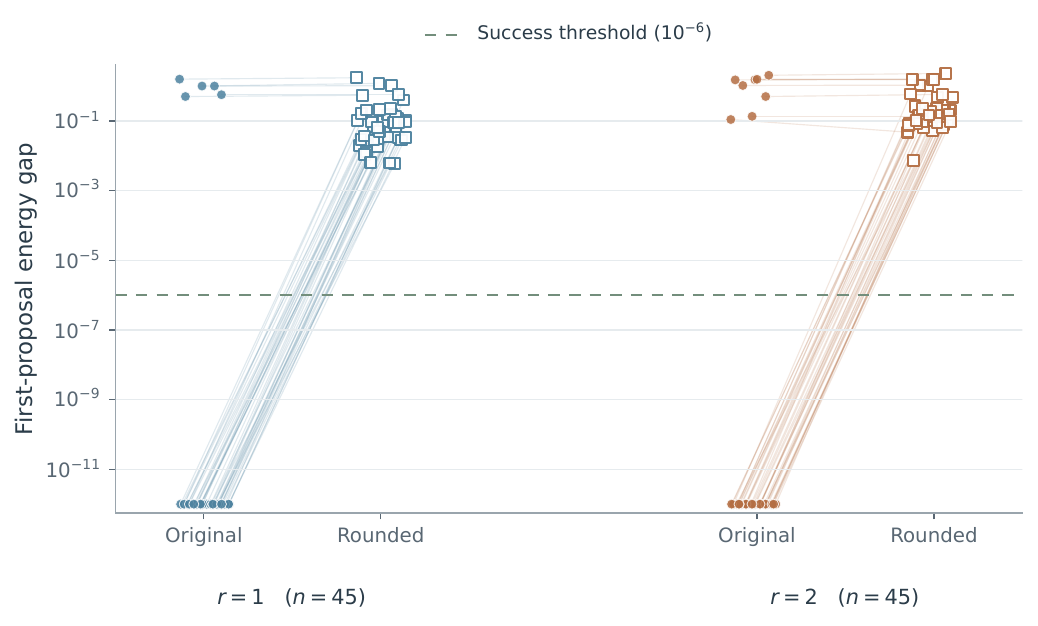}
        \caption{}
        \label{fig:parent_clifford_ablation}
    \end{subfigure}
    \hfill
    \begin{subfigure}[t]{0.49\textwidth}
        \centering
        \includegraphics[width=\linewidth]
        {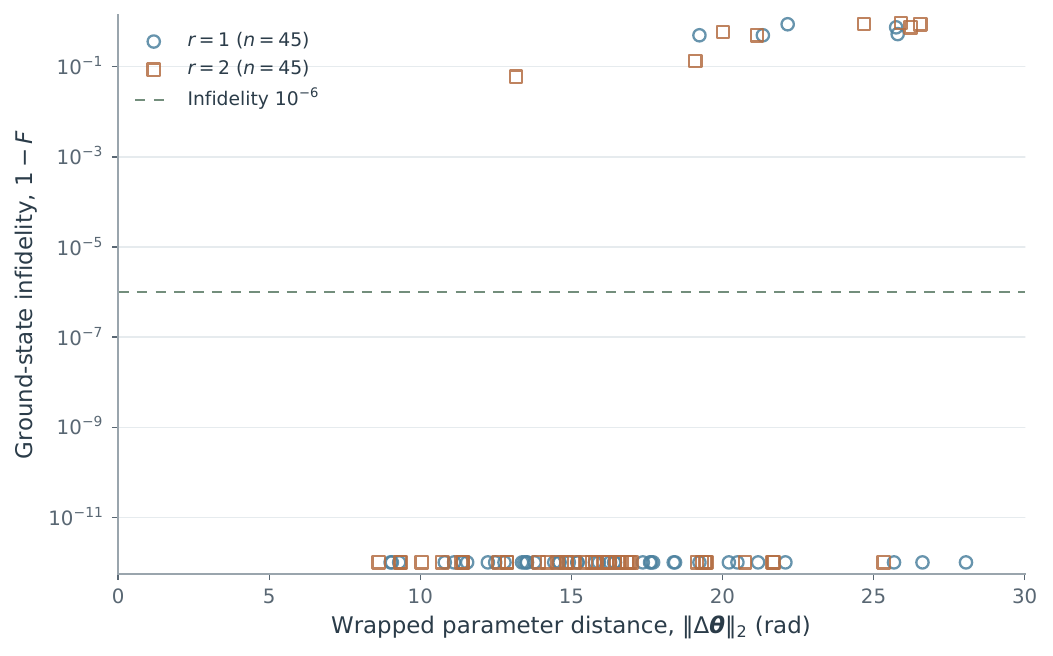}
        \caption{}
        \label{fig:parent_parameter_nonuniqueness}
    \end{subfigure}
    \caption{\justifying
        Structural analysis of the first LMO
        proposals for random parent Hamiltonians.
        (a) Energy gaps before and after rounding
        the proposed angles to Clifford-compatible
        values. The dashed line indicates the
        success threshold of $10^{-6}$.
        (b) Ground-state infidelity versus wrapped
        Euclidean distance from the hidden generating
        parameters. Small infidelities can occur
        despite large parameter distances,
        illustrating the non-uniqueness of
        successful state-preparation parameters.
    }
    \label{fig:parent_structural_analysis}
\end{figure*}

The parent-Hamiltonian instances are generated from
parameter vectors containing predominantly
Clifford-compatible angles, with either one or two
non-Clifford rotations. To assess the contribution
of continuous rotation angles to the LMO proposals,
we perform an ablation in which every angle in the
first proposal is replaced by its nearest
Clifford-compatible value,
\begin{equation}
    \widetilde{\theta}_j
    =
    \operatorname*{argmin}_{c\in\mathcal{C}}
    d_{2\pi}(\theta_j,c),
    \qquad
    \mathcal{C}=\{0,\pm\pi/2,\pi\},
    \label{eq:parent_clifford_rounding}
\end{equation}
where $d_{2\pi}$ denotes angular distance modulo
$2\pi$. The rounded parameter vectors are evaluated
using the original Hamiltonians. Figure~\ref{fig:parent_structural_analysis}(a)
compares the energy gaps before and after rounding.
The original first proposals reach the success
threshold of $10^{-6}$ in 40 of 45 instances for
$r=1$ and 37 of 45 instances for $r=2$.
After rounding, none of the 90 proposals satisfies
this threshold. The median energy increase is
approximately $0.052$ for $r=1$ and $0.109$
for $r=2$. Thus, although the proposed parameter
vectors contain predominantly Clifford-compatible
angles, rounding all angles to this discrete set
substantially reduces their accuracy. We also investigate whether successful proposals
recover the hidden parameter vectors used to
construct the parent Hamiltonians. For each
first proposal, we compute the wrapped Euclidean
distance from the generating parameters,
\begin{equation}
    d_{\theta}
    =
    \left[
    \sum_{j=1}^{p}
    \operatorname{wrap}_{2\pi}
    (\theta_j-\theta_j^{*})^2
    \right]^{1/2},
    \label{eq:parent_parameter_distance}
\end{equation}
and the fidelity with the exact ground state,
\begin{equation}
    F
    =
    \left|
    \braket{\psi^{*}|\psi(\boldsymbol{\theta})}
    \right|^2.
    \label{eq:parent_structural_fidelity}
\end{equation}
The hidden generating parameters are used only
for this post-hoc analysis and are never
provided to the LMO. Figure~\ref{fig:parent_structural_analysis}(b)
shows the ground-state infidelity as a function
of parameter distance. The median wrapped
distances are approximately $15.90$ and
$16.18$ radians for $r=1$ and $r=2$,
respectively, despite median ground-state
fidelities close to unity. Furthermore,
the complete sets of non-Clifford parameter
positions match those of the generating
circuits in only three instances for $r=1$
and two for $r=2$. These results indicate that successful
ground-state preparation does not generally
require recovery of the generating parameters.
Distinct parameter vectors can prepare the
same state without implementing identical
unitaries on arbitrary inputs. Moreover,
the rounding ablation evaluates the
simultaneous replacement of all angles;
it does not establish that each individual
non-Clifford rotation is necessary.

\subsection{Analysis of LMO-Proposed Mechanisms}

In addition to a parameter vector, the LMO
generates a textual explanation describing a
possible physical mechanism underlying its
proposal. We examine these explanations to
characterize their physical content and assess
the consistency of selected claims with
independent numerical calculations.

We analyze the explanations accompanying the
first proposals for all 90 parent-Hamiltonian
instances. Every proposal includes a nonempty
explanation, with median lengths of 42 words
for $r=1$ and 44 words for $r=2$.
Automated textual annotations identify
references to specific Pauli operators in
39 explanations, correlation-related language
in 42, and energy-related language in 86.
These counts describe textual content and
do not constitute assessments of physical
correctness.

The explanations include proposed strategies
based on Pauli eigenstate preparation,
alignment of local rotations with Hamiltonian
terms, and the use of entangling gates to
establish quantum correlations. Such statements
can contain testable claims about intermediate
states, Pauli expectation values, or the
resulting energy. We illustrate their physical
consistency using two contrasting first-proposal
examples, shown in
Figure~\ref{fig:parent_mechanism_examples}.

In the first example, the LMO proposes preparing
$X$-eigenstates such that the fixed $R_{XX}$
gates act only through phases, followed by
a local rotation aligning a qubit with an
effective Hamiltonian field. The checked
state-preparation and field-alignment claims
are consistent with the proposed parameters,
and the resulting energy gap is approximately
zero. In the second example, the LMO states
that its proposed parameters satisfy all
12 commuting parent-Hamiltonian constraints,
predicting zero energy. Independent evaluation,
however, yields an energy gap of $0.5$ and
ground-state fidelity of approximately $0.5$,
contradicting that prediction.

\begin{figure*}[t]
\centering

\begin{tcolorbox}[
    enhanced,
    width=0.95\textwidth,
    boxrule=0.5pt,
    arc=2pt,
    left=8pt,
    right=8pt,
    top=7pt,
    bottom=7pt,
    title={(a) Physically supported proposal
    \hfill $n=4$, $L=1$, $r=1$, seed 42},
    colbacktitle=teal!65!black,
    coltitle=white,
    colframe=teal!65!black,
    colback=white,
    fonttitle=\small\bfseries,
]

\texttt{\small "theta": [24 values omitted]}

\medskip

\texttt{\small "mechanism":}

\begin{quote}
\small\ttfamily
Initialize all qubits in +X, making both RXX gates
global phases. Final rotations align q0 opposite
(0.4598856, 0.1962276, 0), retain q1 and q2 in +X,
and put q3 in +Z. This attains the Hamiltonian
lower bound, approximately zero.
\end{quote}

\small
\textbf{Independent verification.}
$\langle X_0\rangle=-0.919771$,
$\langle Y_0\rangle=-0.392455$, and
$\Delta E\simeq 0$.
The checked physical claims are supported.

\end{tcolorbox}

\vspace{0.5em}

\begin{tcolorbox}[
    enhanced,
    width=0.95\textwidth,
    colbacktitle=teal!65!black,
    coltitle=white,
    colframe=teal!65!black,
    colback=white,
    fonttitle=\small\bfseries,
    boxrule=0.5pt,
    arc=2pt,
    left=8pt,
    right=8pt,
    top=7pt,
    bottom=7pt,
    title={(b) Contradicted proposal
    \hfill $n=12$, $L=2$, $r=2$, seed 42},
    fonttitle=\small\bfseries
]

\texttt{\small "theta": [108 values omitted]}

\medskip

\texttt{\small "mechanism":}

\begin{quote}
\small\ttfamily
Constructs a locally rotated stabilizer state
satisfying all 12 commuting Hamiltonian constraints.
The non-Clifford rotations on q2 and q7 match the
two weighted Pauli pairs. Predicted expectation is
zero up to numerical precision.
\end{quote}

\small
\textbf{Independent verification.}
$\Delta E=0.5$ and $F\simeq 0.5$.
The zero-energy prediction is contradicted.

\end{tcolorbox}

\caption{\justifying
Contrasting examples of LMO-generated physical explanations
for parent-Hamiltonian optimization.
(a) A successful proposal whose checked physical claims are
consistent with independent numerical verification.
(b) A proposal whose predicted zero-energy state is contradicted
by independent evaluation.
The JSON-style excerpts are abridged for presentation:
parameter arrays are omitted and long text is line-wrapped.
}
\label{fig:parent_mechanism_examples}

\end{figure*}

These examples illustrate that model-generated
explanations can contain physically verifiable
claims, but their correctness is not guaranteed.
The two cases are selected to demonstrate
contrasting outcomes and do not provide an
estimate of explanatory correctness across
the full benchmark. Furthermore, agreement
between an explanation and independent
calculations does not establish that the
explanation faithfully represents the model's
internal decision process. Parameter-generation
performance and explanatory correctness
are therefore treated as distinct properties.

\section{Circuit Interpretability: Experimental Protocol and Additional Analysis}
\label{app:circuit_interpretability}

This appendix provides additional details on the circuit
construction, representation transformations, and evaluation
protocol used in the blind circuit-interpretation experiments.
We distinguish the recognition of an algorithmic family from
the prediction of its computational behavior and the physical
consistency of the accompanying explanations.

\subsection{Circuit Construction and Representation Transformations}
\label{app:interpretability_construction}

We consider three algorithmic families: Grover's search,
the quantum Fourier transform (QFT), and
Bernstein-Vazirani. Five instances are generated for
each family, with different problem parameters or input
states. Each instance is represented in logical, native,
and obfuscated forms, resulting in 45 circuit descriptions. For Grover's search, we use three search qubits and
two Grover iterations, with five distinct marked bitstrings.
The resulting measurement distribution favors the marked
state but is not deterministic. For Bernstein-Vazirani,
we use four input qubits and one ancilla, with five distinct
nonzero secret bitstrings. The circuit implements a binary
parity oracle, and the measured input register returns
the corresponding secret in the ideal setting. For QFT, we consider four-qubit circuits with five
independently generated product-state inputs. Each input
is prepared using single-qubit $R_Y$ and $R_Z$ rotations,
followed by the QFT and computational-basis measurements.
These input states provide distinct measurement
distributions and allow the numerical behavior of each
circuit to be evaluated independently.

The logical representation uses standard quantum gates,
with algorithm-specific identifiers and comments removed.
The native representation is obtained by transpiling
the same circuit into the restricted gate set
\begin{equation}
    \mathcal{G}_{\mathrm{native}}
    =
    \{R_Z,R,R_{XX}\},
    \label{eq:interpretability_native_basis}
\end{equation}
where $R$ denotes the parameterized single-qubit
rotation supported by the chosen compilation basis.
The native representation preserves the implemented
computation while replacing familiar high-level
gate sequences with lower-level operations.

To further modify the circuit representation, we
introduce randomized local Clifford frames between
groups of logical gates. Let the original unitary
be decomposed into $m$ consecutive blocks,
\begin{equation}
    U=G_mG_{m-1}\cdots G_1.
\end{equation}
We sample local Clifford transformations
$C_1,\ldots,C_{m-1}$, with boundary conditions
$C_0=C_m=I$, and define
\begin{equation}
    \widetilde{G}_i
    =
    C_iG_iC_{i-1}^{\dagger}.
    \label{eq:interpretability_frame}
\end{equation}
The transformed circuit satisfies
\begin{equation}
    \widetilde{U}
    =
    \widetilde{G}_m\cdots\widetilde{G}_1
    =
    G_m\cdots G_1
    =
    U,
    \label{eq:interpretability_frame_equivalence}
\end{equation}
because the intermediate frame transformations
cancel between consecutive blocks. In our implementation, frames are sampled from
tensor products of single-qubit Clifford gates,
with transformations introduced between groups
of four logical operations. Each transformed
block is independently compiled into
$\mathcal{G}_{\mathrm{native}}$.
This blockwise compilation avoids applying a
final global optimization that could eliminate
many of the inserted frame transformations.
The procedure changes the gate-level description
without changing the implemented unitary. 

The logical, native, and obfuscated representations
are generated from the same underlying instance.
Only the resulting OpenQASM 3 description is
provided to the LLM. The original circuit,
transformation metadata, expected outcomes,
and equivalence checks are excluded from
the inference prompt. The randomized-frame construction is used as a controlled representation transformation rather
than a cryptographic obfuscation method. In
particular, the transformed circuits may contain
additional gates and longer descriptions, and
their structure may remain recoverable through
suitable circuit simplification.

\subsection{Interpretation Prompt and Evaluation Protocol}
\label{app:interpretability_evaluation}

All circuit interpretations are generated using
the same frozen language model and fixed
system instructions. The model receives only
the anonymized OpenQASM 3 description of an
instantiated circuit and is instructed to
analyze its operations without using external
tools, simulations, or experimental feedback. The requested response is a structured JSON
object containing a description of the
circuit's mathematical behavior, a hypothesis
about its computational purpose, predicted
measurement behavior, supporting gate-level
evidence, alternative interpretations,
properties that cannot be identified, and
a confidence value associated with the
algorithm hypothesis. The inference protocol also permits explicit
predictions of the most likely output bitstring
and, when feasible, a measurement probability
distribution. The model may abstain from these
numerical predictions when they cannot be
reliably inferred. No ground-truth information
is provided during inference, and all
interpretations are generated without feedback
from previous circuit evaluations.

We distinguish three aspects of the
generated interpretations. First, algorithm identification evaluates
whether the reported algorithm hypothesis
matches the known circuit family. This
criterion measures recognition of the
algorithmic structure but does not establish
correctness of the predicted computation. Second, functional output prediction evaluates the explicit numerical predictions
against independently calculated outcomes.
For Grover and Bernstein-Vazirani, we
compare the predicted most-likely bitstring
with the exact most-likely measurement
outcome. A correct prediction receives
a score of one; an incorrect or absent
prediction receives zero when calculating
accuracy over all valid responses. Third, physical explanation consistency concerns the supporting claims produced by
the model. These may describe the roles of
specific gates, interference, phase
accumulation, or the relationship between
the circuit structure and its computational
output. Such explanations are evaluated
separately from algorithm identification
and numerical output prediction.

\subsection{Additional Analysis of Model-Generated Interpretations}
\label{app:interpretability_additional}

The complete benchmark contains 45 valid
interpretations, with five instances for
each combination of algorithm and circuit
representation. All three representations
are evaluated independently, allowing
algorithm identification and functional
prediction to be compared for the same
underlying computations. The results reveal a distinction between algorithm identification and numerical
output prediction. In the logical
representation, the model identifies
all 15 algorithmic instances. Under
native decomposition and obfuscation,
the corresponding totals are 11 and
12 correct identifications, respectively.

This distinction is particularly pronounced
for Bernstein-Vazirani. Although the
model identifies the algorithm in only
two of five native and two of five
obfuscated instances, it predicts the
correct secret bitstring in all ten.
Thus, the numerical output can remain
correct even when the conventional
algorithmic label is not recovered. For Grover's algorithm, the model correctly predicts the most likely output in five
logical, four native, and five obfuscated
instances. For QFT, the algorithm is
identified in all 15 cases. However,
these identification results alone
do not determine whether the model's
explanations correctly describe the
underlying quantum transformations.

%
%

%
%

The transformations also modify circuit
depth and OpenQASM description length.
Consequently, differences across
representations may reflect both changes
in recognizable gate patterns and the
additional complexity of the circuit
descriptions. The experiment does not
isolate these effects or establish that
the model's interpretations are independent
of previously learned algorithmic patterns. Nevertheless, the results demonstrate
that algorithm identification and functional
output prediction need not vary together
under equivalent circuit representations.
This distinction motivates evaluating
the mathematical behavior inferred by
a language model separately from its
ability to recognize familiar quantum
algorithm structures.

Boxes~\ref{box:interpretation_bv_logical}
and~\ref{box:interpretation_bv_obfuscated}
present the complete LLM responses for two equivalent
representations of the same Bernstein--Vazirani
instance. In the logical representation, the model
identifies Bernstein--Vazirani as its most plausible
interpretation and correctly predicts the output
bitstring $0100$. In the obfuscated representation,
the model does not identify a unique algorithmic
purpose but still predicts the same output,
illustrating the distinction between algorithm
identification and functional prediction.

\begin{interpretjson}[label=box:interpretation_bv_logical]{(a) Logical representation}
{
  "circuit_behavior": "The circuit prepares q[0] through q[3] in |+> and q[4] in |->. The CNOT implements the oracle f(x)=x[2]. Because X|->=-|->, it introduces the phase (-1)^x[2] without entangling the ancilla with the input register. The final Hadamards leave q[0], q[1], and q[3] in |0>, and q[2] in |1>. The unmeasured ancilla remains in |->.",
  "algorithm_hypothesis": "Bernstein\u2013Vazirani with hidden string 0100 is the best-fitting interpretation, but the intended task is ambiguous.",
  "predicted_measurement_behavior": "In ideal execution, the measurements always give c[3]=0, c[2]=1, c[1]=0, and c[0]=0.",
  "predicted_most_likely_bitstring": "0100",
  "predicted_probabilities": {
    "0100": 1.0
  },
  "evidence": [
    "The X followed by H on q[4] prepares the phase-kickback ancilla |->.",
    "Initial Hadamards prepare an equal superposition of all 16 input-register basis states.",
    "The only oracle gate is cx q[2], q[4], corresponding to the linear Boolean function f(x)=x[2].",
    "Phase kickback changes q[2] from |+> to |->; the final Hadamard maps it to |1>.",
    "On q[0], q[1], and q[3], the two Hadamards cancel, leaving |0>.",
    "The measurement mapping places the sole 1 in c[2], giving 0100 in descending classical-bit order."
  ],
  "alternative_hypotheses": [
    "A Deutsch\u2013Jozsa circuit for the balanced function f(x)=x[2]; its nonzero output certifies the balanced case under the usual promise.",
    "A demonstration of phase kickback or Hadamard-basis interference without an intended oracle-identification task."
  ],
  "not_identifiable": [
    "Whether the intended task is hidden-string recovery, constant-versus-balanced classification, or a gate-level demonstration.",
    "Whether the oracle is intended to be unknown to an executing algorithm.",
    "Physical noise rates and actual hardware outcome frequencies."
  ],
  "confidence": 0.85
}
\end{interpretjson}
\noindent{\small\textbf{Independent verification.} Exact most-likely output: \texttt{0100} ($P=1$). The model identifies Bernstein-Vazirani as its preferred interpretation.}\par
\medskip

\clearpage
\begin{interpretjson}[label=box:interpretation_bv_obfuscated]{(b) Obfuscated representation}
{
  "circuit_behavior": "Starting from all-zero qubits, the circuit produces |0> on q[0], |0> on q[1], |1> on q[2], |0> on q[3], and |-> on q[4], up to an overall phase. The sole two-qubit interaction does not entangle this input: immediately before it, q[2] is |1> and q[4] is the X eigenstate |->. Consequently, rxx(pi/2) acts as Rx(-pi/2) on q[2]. Subsequent rotations restore q[2] to |1> and leave q[4] in |-> up to phase.",
  "algorithm_hypothesis": "not identifiable",
  "predicted_measurement_behavior": "In ideal execution, every shot returns c[3]=0, c[2]=1, c[1]=0, c[0]=0. The unmeasured fifth qubit does not introduce randomness into these results.",
  "predicted_most_likely_bitstring": "0100",
  "predicted_probabilities": {
    "0100": 1.0
  },
  "evidence": [
    "The custom gate r(theta, phi) equals exp[-i theta (cos(phi) X + sin(phi) Y)/2]. Thus r(theta, 0)=Rx(theta) and r(theta, pi/2)=Ry(theta).",
    "A chronological pair Ry(pi/2), Rx(pi) implements -iH. On q[0], four such pairs and two additional Rx(pi) gates reduce to identity up to phase. On q[1], four such pairs also reduce to identity.",
    "On q[3], two such pairs followed by two Rz(pi) rotations reduce to identity up to phase.",
    "Before rxx, the opposite Rz(pi/2) rotations on q[2] cancel, as do its two adjacent Rz(pi) rotations up to phase. Its remaining Ry(pi/2), Rx(pi), Ry(pi/2) sequence equals Rx(pi), preparing |1>.",
    "Before rxx, the Rz(pi) rotations on q[4] cancel in pairs up to phase. Its remaining Rx(pi), Ry(pi/2), Rx(pi) sequence prepares |-> up to phase.",
    "The supplied rxx decomposition implements exp[-i theta X\u2297X/2]. Since X|->=-|->, its action here is Rx(-pi/2) on q[2].",
    "After rxx, the opposite Y rotations on q[2] cancel. Combining the interaction's effective Rx(-pi/2), the explicit Rx(-pi/2), and the final Rx(pi) gives identity. The final Rx(-pi/2) on q[4] contributes only a phase."
  ],
  "alternative_hypotheses": [
    "A deterministic state-preparation circuit containing redundant rotations.",
    "A gate-rewriting or compilation test whose intended significance is not visible in its final measurement statistics."
  ],
  "not_identifiable": [
    "A unique named algorithm or intended computational problem.",
    "Whether the redundant gates are intentional, compiler-generated, or part of a larger omitted computation."
  ],
  "confidence": 0.1
}
\end{interpretjson}
\noindent{\small\textbf{Independent verification.} Exact most-likely output: \texttt{0100} ($P=1$). The model does not identify a unique algorithmic purpose.}

\end{document}